\documentclass[acmsmall,screen]{acmart}
\usepackage{xcolor}
\usepackage{pifont}
\usepackage{booktabs}
\usepackage{tabularx}
\usepackage{graphicx}
\usepackage{tcolorbox}
\usepackage{multirow}
\usepackage{enumitem}
\usepackage{colortbl}
\usepackage{subcaption} 
\usepackage{soul}
\usepackage{threeparttable}
\AtBeginDocument{%
  }

\setcopyright{acmlicensed}
\copyrightyear{2026}
\acmYear{2026}
\acmDOI{XXXXXXX.XXXXXXX}

\acmISBN{978-1-4503-XXXX-X/18/06}

\newcommand{\find}[1]{
\begin{tcolorbox}[leftrule=1mm,toprule=0mm,bottomrule=0mm,left=1pt,right=2pt,top=2pt,bottom=2pt]
\em #1
\end{tcolorbox}
}

\definecolor{dimgreen}{RGB}{34, 139, 34} 

\begin{document}

\setcounter{page}{1}
\title[\textsc{Mapping NVD Records to Their VFCs}]{Mapping NVD Records to Their Vulnerability-fixing Commits: How Hard is It?}

\author{Huu Hung Nguyen}
\affiliation{%
  \institution{Singapore Management University}
  \country{Singapore}
}
\email{huuhungn@smu.edu.sg}

\author{Ting Zhang}
\affiliation{%
 \institution{Monash University}
 \country{Australia}}
\email{ting.zhang@monash.edu}
\authornote{Ting Zhang is the corresponding author.}

\author{Duc Manh Tran}
\email{dmtran@smu.edu.sg}
\affiliation{%
  \institution{Singapore Management University}
  \country{Singapore}
}

\author{Yiran Cheng}
\email{chengyiran@iie.ac.cn}
\affiliation{%
  \institution{University of Chinese Academy of Sciences}
  \country{China}
}

\author{Thanh Le-Cong}
\email{congthanh_le@sutd.edu.sg}
\affiliation{%
  \institution{Singapore University of Technology and Design}
  \country{Singapore}
}

\author{Hong Jin Kang}
\affiliation{%
  \institution{University of Sydney}
  \country{Australia}}
\email{hongjin.kang@sydney.edu.au}

\author{Ratnadira Widyasari}
\email{ratnadiraw@smu.edu.sg}
\affiliation{%
  \institution{Singapore Management University}
  \country{Singapore}
}

\author{Shar Lwin Khin}
\affiliation{
\institution{Singapore Management University}
\country{Singapore}
}
\email{lkshar@smu.edu.sg}

\author{Ouh Eng Lieh}
\affiliation{
\institution{Singapore Management University}
\country{Singapore}
}
\email{elouh@smu.edu.sg}

\author{David Lo}
\affiliation{%
  \institution{Singapore Management University}
  \country{Singapore}
}
\email{davidlo@smu.edu.sg}

\renewcommand{\shortauthors}{Nguyen et al.}
\begin{abstract}
Mapping National Vulnerability Database (NVD) records to vulnerability-fixing commits (VFCs) is crucial for vulnerability analysis but challenging due to sparse explicit links in NVD references. This study explores this mapping's feasibility through an empirical approach. Manual analysis of NVD references showed Git references enable over 86\% success, while non-Git references achieve under 14\%. Using these findings, we built an automated pipeline extracting 31,942 VFCs from 20,360 NVD records (8.7\% of 235,341) with 87\% precision, mainly from Git references. To fill gaps, we mined six external security databases, yielding 29,254 VFCs for 18,985 records (8.1\%) at 88.4\% precision, and GitHub repositories, adding 3,686 VFCs for 2,795 records (1.2\%) at 73\% precision. Combining these, we mapped 26,710 unique records (11.3\% coverage) from 7,634 projects, with overlap between NVD and external databases, plus unique GitHub contributions. Despite success with Git references, 88.7\% of records remain unmapped, highlighting the difficulty without Git links. This study offers insights for enhancing vulnerability datasets and guiding future automated security research.

% Mapping National Vulnerability Database (NVD) records to vulnerability-fixing commits (VFCs) is vital for understanding vulnerability fixes and data-driven vulnerability analysis; yet, it remains inherently difficult due to sparse explicit VFC links in NVD references.

% This study investigates the feasibility and difficulty of this mapping through a comprehensive empirical approach.
% We began with a manual analysis of NVD references across four record categories, revealing that Git references enable high success (>86\%), while non-Git references yield under 14\%. 
% Leveraging these insights, we developed an automated pipeline that extracted 31,942 VFCs from 20,360 NVD records (8.7\% of 235,341 total records) with 87\% precision, primarily from Git references. 
% To address gaps, we mined six external security databases, identifying 29,254 VFCs for 18,985 records (8.1\%) at 88.4\% precision, and analyzed GitHub repositories, adding 3,686 VFCs for 2,795 records (1.2\%) at 73\% precision. 
% Combining these sources, we mapped 26,710 unique records (11.3\% coverage) from 7,634 projects, with significant overlap between NVD and external databases, and unique contributions from GitHub. 

% Our findings highlight that while Git references are reliable, 88.7\% of NVD records remain unmapped, underscoring the persistent difficulty of linking records without Git references to their VFCs.
% This work provides actionable insights for improving vulnerability datasets and informs future research in automated security analysis.
\end{abstract}




\maketitle

\section{Introduction}
\label{sec:introduction}

% vulnerability is severe
Open-source software (OSS) vulnerabilities are a prime target for malicious actors, often leading to severe security breaches, including data leaks and system disruptions~\cite{o2019privilege,biswas2018study,cheng2017enterprise,20183,turner2008symantec}. To standardize the tracking of these vulnerabilities, the Common Vulnerabilities and Exposures (CVE) system~\cite{cve:online}, managed by MITRE, assigns unique identifiers to each disclosed vulnerability. These identifiers are further enriched by the National Institute of Standards and Technology (NIST) in the National Vulnerability Database (NVD), which provides additional details such as severity ratings and references to patches or fixes~\cite{NVDHome30:online}. With over 200,000 records, the NVD serves as a cornerstone for vulnerability research. However, while the NVD offers a wealth of information, it often falls short of providing a complete picture of the vulnerability lifecycle, particularly in understanding how vulnerabilities are mitigated in practice.

% \begin{figure*}[ht]
% \centering
% \includegraphics[width=0.7\textwidth]{figure/nvd_exp.png}
% \caption{Partial screenshot of the NVD record for CVE-2023-33976}
% \label{fig:nvd_exp}
% \end{figure*}

A critical aspect of this lifecycle is the identification and analysis of vulnerability-fixing commits (VFCs), which are essential for understanding how vulnerabilities are resolved and tracing the commits that introduced them in the first place~\cite{le2021deepcva,perl2015vccfinder}. VFCs, often referred to as security patches, are not only crucial for resolving vulnerabilities but also for building comprehensive datasets that can inform future research and development. While some NVD records include Git links to VFCs in their reference fields, marked with the \texttt{Patch} tag, such links are scarce; only 6.7\% of NVD records provide this information. This scarcity highlights a significant gap in the availability of actionable data for researchers and practitioners.

The study of historical VFCs offers valuable insights into vulnerability trends, recurring patterns, and the broader context of how vulnerabilities emerge across diverse software systems~\cite{scan2014historical,mell2007complete,dunlap2024vfcfinder}. 
Moreover, detailed information on VFCs is indispensable for advancing automated vulnerability detection methods, especially as deep learning and data-driven approaches become increasingly popular in security research~\cite{nikitopoulos2021crossvul,bhandari2021cvefixes,prospector,dunlap2024vfcfinder}. However, the limited availability of VFCs poses a significant challenge, hindering progress in both research and practical applications.

A critical research problem arises in mapping NVD records to their corresponding VFCs. While prior work has explored various methods for identifying VFCs~\cite{bhandari2021cvefixes,prospector,li2024patchfinder,dunlap2024vfcfinder,wang2022vcmatch}, the focus has largely been on improving the accuracy of general VFC identification rather than specifically mapping NVD records to their fixes. Recent studies have attempted to extract VFC information from NVD records~\cite{nikitopoulos2021crossvul,bhandari2021cvefixes,bigvul} or proposed methods to map NVD records to their VFCs~\cite{li2024patchfinder,prospector}. 
However, a critical gap remains in the literature: there has been no large-scale, systematic investigation into the feasibility and inherent difficulty of comprehensively mapping NVD records to their VFCs.
This gap drives our research, as understanding the difficulty of this mapping is crucial to unlocking the NVD's full potential for vulnerability analysis.
This research problem extends the scope of prior work by not only aiming to enhance the accuracy of VFC identification but also seeking to maximize the coverage of identified mappings. 
Despite its importance, there is a lack of systematic investigation into the challenges of this process. To fill this gap, our study explores the difficulty of identifying NVD-VFC mappings and provides actionable insights for future research in software security. Specifically, we address the following research questions (RQs):

\begin{itemize}[leftmargin=0pt, itemindent=*]
    \item \textbf{RQ1:} What methods do humans use to identify VFCs associated with a given NVD record?
    \item \textbf{RQ2:} What proportion of records in the NVD include patch links in their reference sections?
    \item \textbf{RQ3:} How many patch links can be discovered from external security databases?
    \item \textbf{RQ4:} How many patch links can be identified through GitHub repository analysis?
    \item \textbf{RQ5:} How do the VFCs extracted from NVD references, external security databases, and GitHub repositories compare in terms of precision, coverage, and overlap?
\end{itemize}

In this paper, we present an empirical study to tackle this challenge head-on, exploring the feasibility and obstacles of linking NVD records to VFCs across their diverse categories. Unlike prior work that often focuses on individual methods in isolation, our study is the first to systematically combine and compare multiple VFC identification approaches in a unified framework. We began with a manual investigation of NVD references, categorizing records by Git and Patch-tag presence to assess human identification strategies and their success rates. Building on these findings, we developed an automated pipeline to extract VFCs from NVD references, achieving high precision for Git-linked records. To address remaining gaps, we mined external security databases (e.g., Snyk, GitHub Advisory) and analyzed GitHub repositories using state-of-the-art tools like Prospector~\cite{prospector}, comparing all sources for precision, coverage, and overlap. This allows us to provide the first systematic comparison of the precision, coverage, and overlap across all sources.
Our results show that while 26,710 records (11.3\% of the NVD) can be mapped with 86.1\% VFC precision, around 88.7\% remain unmapped, particularly those without Git references, highlighting significant challenges and opportunities.
% \thanh{This paragraph said that 13\% NVD mapped but only 85\% unmapped so where is another 2\%? @Hung: You may want to double check the numbers or provide explainations}
% \zt{we said over 85\% here. maybe we will just say 88\% to avoid misunderstanding}

Our contributions can be summarized as follows:
% \vspace{-10pt}

\begin{itemize}[leftmargin=0pt, itemindent=*]
\item We performed a manual investigation to understand how humans identify VFCs from NVD references.
\item We performed the first comprehensive empirical study that systematically integrates and compares VFCs from three primary sources: NVD references, external security databases, and GitHub repositories.
\item We evaluated the effectiveness of extracting VFCs from these three sources, comparing precision, coverage, and overlap to offer practical insights for future research and applications.
\item We compiled a dataset of 37,441 VFCs across 26,710 NVD records, achieving an average precision of 86.1\% - to our knowledge, the largest VFC dataset to date.
\end{itemize}

It is important to note that the primary contribution of this work is a large-scale, comprehensive empirical study. While the VFC dataset is a key outcome, our main novelty lies in the design and execution of the first holistic investigation into the NVD-VFC mapping problem. We systematically explore, quantify, and compare multiple mapping strategies at a scale not previously attempted. This provides the first concrete evidence for the "hardness" of this task and establishes a crucial baseline for all future work aiming to develop fully automated solutions.

The rest of this article is organized as follows: Section~\ref{sec:background} discusses the background; 
Section~\ref{sec:study_design} presents our study design; 
Section~\ref{sec:results} presents the experimental results; Section~\ref{sec:discussion} elaborates on the lessons learned and threats to validity; 
Section~\ref{sec:related_work} discusses other relevant studies;
finally, Section~\ref{sec:conclusion} concludes our study and outlines potential future work.

\section{Background and Related Work}
\label{sec:background}
In this section, we provide the background knowledge related to our work. 
First, we provide an overview of VFCs. 
Second, we review existing approaches to VFC mapping and their limitations.
Third, we discuss the challenges associated with mapping NVD records to their VFCs. 
Last, we demonstrate the different NVD record categories.

\subsection{Overview of Vulnerability-fixing Commits}
VFCs, also known as \textit{security patches}, are code changes that fix or mitigate vulnerabilities. 
VFCs play a crucial role in various vulnerability-related research areas, such as vulnerability type detection~\cite{pan2023fine}, vulnerability-introducing commit detection~\cite{bao2022v}, and automated program repair~\cite{fu2022vulrepair}. 
They enable researchers to perform in-depth analyses of vulnerabilities and the corresponding code changes, providing valuable insights into how vulnerabilities are introduced and resolved.

Moreover, VFCs are essential for developing automated approaches to vulnerability detection and repair, and vulnerable OSS version identification~\cite{pan2023fine,fu2022vulrepair,cheng2024llm}. 
Despite the NVD publicly disclosing numerous vulnerabilities, a significant portion of vulnerabilities and their corresponding VFCs remain undisclosed~\cite{lin2024vulnerabilities}. 
This gap in disclosure poses challenges for both security research and practical applications.

\subsection{Existing Approaches to VFC Mapping}
\label{sec:mapping}
Most studies identify VFCs by directly utilizing Git links provided in NVD references~\cite{nikitopoulos2021crossvul, bigvul, bhandari2021cvefixes, nguyen2022vulcurator, akhoundali2024morefixes}. For example, the CrossVul dataset~\cite{nikitopoulos2021crossvul} collects GitHub commits from NVD references and processes pull request references to obtain all relevant commits. Similarly, both the Big-Vul~\cite{bigvul} and CVEFixes~\cite{bhandari2021cvefixes} datasets rely on reference links to Git repositories for VFCs, extending beyond GitHub to include repositories like Google Android (Big-Vul) and GitLab and Bitbucket (CVEFixes).

Other studies extract CVE and VFC mappings from external security databases. For instance, the DiverseVul dataset~\cite{chen2023diversevul} uses databases from Snyk.io and bugzilla.redhat.com, while Sun et al.~\cite{sun2023silent} leverage the GitHub Advisory Database.

Another approach involves searching GitHub repositories directly. Although this area has recently gained attention, few studies exist~\cite{wang2022vcmatch, xu2022tracking, hommersom2024automated, prospector, li2024patchfinder}. Notable recent methods include FixFinder~\cite{hommersom2024automated}, which preceded Prospector~\cite{prospector}, and PatchFinder~\cite{li2024patchfinder}, which has demonstrated improved performance over earlier methods like PatchScout~\cite{patchscout} and VCMatch~\cite{wang2022vcmatch}.
MoreFixes~\cite{akhoundali2024morefixes}, a recent dataset, builds upon Prospector~\cite{prospector} by extracting GitHub repository URLs from GitHub Security Advisories~\cite{advisories} and NVD references, then employing Prospector to identify and score potential VFCs using the repository URL and CVE-ID.

To provide a quantitative overview of the datasets produced by these various approaches, Table~\ref{tab:dataset_comparison} compares several state-of-the-art VFC datasets.

\begin{table}[h!]
\centering
\caption{Quantitative Comparison of VFC Datasets}
\label{tab:dataset_comparison}
\begin{tabular}{@{}lrrr@{}}
\toprule
\textbf{Dataset} & \textbf{\# Records} & \textbf{\# VFCs} & \textbf{Primary Collection Method(s)}                                                                                   \\ \midrule
BigVul           & 3,754            & 4,432            & NVD Git References                                                                                                      \\
CrossVul         & 5,131            & 5,877            & NVD Git References                                                                                                      \\
CVEFixes         & 5,365            & 5,495            & NVD Git References                                                                                                      \\
MoreFixes        & 26,617           & 31,883           & GitHub Repos Mining                                                                                                     \\
\textbf{Ours}    & \textbf{26,710}  & \textbf{37,441}  & \textbf{\begin{tabular}[c]{@{}r@{}}NVD References (Git \& Non-Git) + \\ External Databases + Repos Mining\end{tabular}} \\ \bottomrule
\end{tabular}%
\end{table}

% \st{Most existing studies focus narrowly on a single aspect of NVD record mapping, failing to provide a comprehensive perspective. 
% In contrast, our study offers the most holistic approach to date for mapping NVD records to VFCs. 
% We begin by integrating a dedicated manual investigation to inform and refine automated extraction processes. 
% Unlike prior works that rely solely on NVD reference links, we enrich our analysis by incorporating data from external security databases and GitHub repositories,  improving the depth of VFC identification. 
% Furthermore, our methodology stands out through multiple rounds of manual verification on sampled VFCs at each stage, ensuring high precision in identifying the relevant commits.}

\noindent While the table shows a progression in dataset size, it also highlights a key limitation of prior work: most studies focus narrowly on a single collection strategy, failing to provide a comprehensive perspective.

In contrast, our study offers the most holistic approach to date. We demonstrate this by systematically integrating and comparing the very methods that prior studies have applied in isolation. For instance, while works like MoreFixes~\cite{akhoundali2024morefixes} and CVEFixes~\cite{bhandari2021cvefixes} extract VFCs from Git references, and DiverseVul~\cite{chen2023diversevul} uses external databases, no prior work has provided a comparative analysis of these sources. Although the MoreFixes dataset is comparable in volume to ours, it relies heavily on repository mining tools (e.g., Prospector), which our empirical evaluation (RQ4) demonstrates suffer from degraded precision (~73\%) when identifying implicit VFCs.
On the contrary, over 90\% of our VFCs are derived from external security databases and curated NVD references, achieving precisions of 88.4\% and 87.0\%, respectively. This high-precision approach minimizes false positives, providing a cleaner ground truth for data-driven tasks like automated vulnerability repair and vulnerability-introducing commit detection, where model accuracy is strictly tied to training data quality.

To summarize, our approach is novel in three ways: (1) We are the first to systematically mine non-Git references in NVD records through our NVD reference tree, a source completely overlooked by existing studies, (2) We integrate multiple extraction methods (NVD references, external databases, and GitHub repositories) in a unified pipeline and empirically compare their effectiveness and (3) We begin with a dedicated manual investigation to establish ground truth and inform our automated processes, unlike prior works that go directly to automation.

Furthermore, our methodology stands out through multiple rounds of manual verification on sampled VFCs at each stage, ensuring high precision in identifying the relevant commits. This combination of new extraction methods (non-Git references), systematic integration of existing approaches, and rigorous empirical comparison constitutes a significant methodological advance beyond simply reusing existing techniques.

% In contrast, our approach follows a more comprehensive process to map NVD records to their VFCs. 
% First, we conducted a manual investigation to understand how human may look for VFCs from NVD records.
% Next, we utilize direct links provided in the NVD references. Specifically, we examine not only Git links but also other types of links to ensure comprehensive coverage.
% Subsequently, we incorporate information from additional external security databases beyond GitHub Security Advisories. For NVD records where VFCs are still not identified, we perform targeted searches within the relevant GitHub repositories.
% At each step of this process, we conduct manual verification to ensure precision before proceeding to the next stage, thereby enhancing the reliability of the extracted VFCs.

\subsection{Challenges in Mapping NVD Records to their VFCs}
Despite the importance of VFCs, mapping NVD records to their corresponding VFCs is challenging due to factors such as the following:

\begin{itemize}[leftmargin=0pt, itemindent=*]
    \item \textbf{Limited Git links in NVD references:} Only a small percentage of NVD records include Git links that are marked with \texttt{Patch} tag, even if all these links are assumed to be true VFCs. For the majority of NVD records, alternative strategies are required, such as analyzing other reference links or seeking external security databases. One potential approach involves searching in relevant GitHub repositories.
    
    \item \textbf{Extensive commit history in OSS repositories:} Even with CPE (Common Platform Enumeration) information from NVD records, identifying relevant VFCs in OSS GitHub repositories is daunting. Large repositories often contain hundreds of thousands of commits, making it challenging to pinpoint the specific commits addressing vulnerabilities.

    \item \textbf{Semantic disparity between NVD descriptions and VFCs:} NVD descriptions have been employed to identify VFCs from a pool of commits~\cite{li2024patchfinder}. However, these descriptions are often verbose, whereas commit messages are typically succinct. Additionally, while code changes in VFCs offer valuable context, aligning them with natural language descriptions poses challenges due to the differing modalities (code vs. text), complicating the matching process.
\end{itemize}

\subsection{NVD Record Categories}
\label{sec:nvd_categories}
To systematically identify VFCs, we categorize NVD records based on the types of links in their reference sections.
Since VFCs are typically associated with Git, the presence or absence of Git links is a critical factor in categorizing NVD records. 
Additionally, as the NVD assigns a \texttt{Patch} tag to certain reference links, whether a reference includes this tag is another important criterion.

To facilitate our manual investigation, we define four mutually exclusive categories of NVD records based on these two factors:

\begin{itemize}[leftmargin=0pt, itemindent=*]
    \item \textbf{Category 1:} NVD records containing at least one Git reference with a \texttt{Patch} tag: Commits associated with these Git references could be considered VFCs with high confidence. 
    
    \item \textbf{Category 2:} NVD records containing at least one Git reference, but none of them have a \texttt{Patch} tag: While these non-patch-tagged Git references are likely to contain VFCs, the NVD has not yet confirmed their validity.

    \item \textbf{Category 3:} NVD records with no Git references but containing at least one other reference with a \texttt{Patch} tag: While these references are highly relevant to VFCs, it remains unclear how to extract VFCs from them.
    
    \item \textbf{Category 4:} NVD records with no Git references and no other references with a \texttt{Patch} tag: These are the remaining NVD records. These cases can be the most challenging among all the NVD records, as they lack any explicit relationship with VFCs and may be entirely irrelevant.
\end{itemize}

Considering these four categories, the challenges are twofold: for NVD records with Git references (Categories 1 and 2), we need to manually verify whether the references correspond to true VFCs. 
Meanwhile, for NVD records with only non-Git references (Category 3 and 4), we need to first follow those links to uncover any potential Git references that could lead to VFCs; then, we also manually verify their precision.

\section{Study Design}
\label{sec:study_design}

In this section, we first introduce the RQs and their motivation. 
Second, we describe the approach to answer each RQ.
Third, we present the data collection process. 
Next, we provide the evaluation metrics adopted in each RQ.
Finally, we provide the implementation details.

\subsection{Research Questions}
\textbf{RQ1: What methods do humans use to identify VFCs associated with a given NVD record?}

\vspace{2px}
\noindent\textbf{\textit{Motivation:}}
Understanding how humans manually identify VFCs from NVD references is foundational for developing automated techniques.
In this RQ, we present a study on manual VFC identification using only the existing reference links in NVD records. 
The goal of this investigation is to identify and categorize the different types of references found in NVD records and assess the potential of each reference type in uncovering VFCs. 
By manually analyzing the samples of NVD records and their associated references, we aim to gain insights into the characteristics and patterns of references that are most likely to lead to the identification of VFCs. 
This manual process will help us establish a set of rules and heuristics that can form the basis for automating the VFC identification process.

\vspace{4px}
\noindent\textbf{RQ2: What proportion of records in the NVD include patch links in their reference sections?}

\vspace{2px}
\noindent\textbf{\textit{Motivation:}} RQ1 revealed that VFCs can be reliably identified from NVD references, especially Git references, regardless of \texttt{Patch} tags. 
To scale this process, we developed an automated pipeline to extract VFCs systematically, aiming to quantify their effectiveness and precision across all NVD record categories.

\vspace{4px}
\noindent\textbf{RQ3: How many patch links can be discovered from external security databases?}

\vspace{2px}
\noindent\textbf{\textit{Motivation:}} While the NVD is a primary source of vulnerability information, it is not exhaustive in its coverage of VFCs. 
Many NVD records lack explicit VFC links or references, which limits their utility for developers and researchers seeking to understand or remediate similar vulnerabilities. 
However, other security databases and platforms often host complementary CVE information, including additional references, patches, or commit links that may not be present in the NVD. 
By exploring these external databases, we aim to uncover additional VFCs that are missing from the NVD records. 

\vspace{4px}
\noindent\textbf{RQ4: How many patch links can be identified through GitHub repository analysis?}

\vspace{2px}
\noindent\textbf{\textit{Motivation:}} Despite our efforts to extract VFCs from the NVD and external security databases, we observed that a significant proportion of NVD records still lack explicit VFC links. 
This gap highlights the need for alternative approaches to identify VFCs. 
GitHub, as one of the largest platforms for hosting open-source projects, is a natural candidate for this exploration. 
Many vulnerabilities are patched directly in GitHub repositories, and these patches are often documented in commits, pull requests, or issues. 
By leveraging GitHub repository analysis, we aim to identify additional VFCs that are not referenced in the NVD or external databases.

\vspace{4px}
\noindent\textbf{RQ5: How do the VFCs extracted from NVD references, external security databases, and GitHub repositories compare in terms of precision, coverage, and overlap?}

\vspace{2px}
\noindent\textbf{\textit{Motivation:}} By employing multiple strategies to identify VFCs - ranging from automated extraction from NVD reference (S1) and external databases (S2) to GitHub repositories (S3) - we have gathered a diverse set of VFCs. 
However, it is unclear how these sources compare in terms of their precision (the correctness of the VFCs identified from each source), coverage (the proportion of NVD records for which VFCs are found), and overlap (the extent to which VFCs from different sources are unique or redundant). 
Understanding these dimensions will also provide insights into the complementary nature of these sources and inform future efforts to automate and optimize VFC discovery.

\subsection{Methodology}
\subsubsection{RQ1: Manual Investigation}
We sampled a representative set of NVD records from each category (as shown in Section~\ref{sec:nvd_categories}) to maintain a 95\% confidence level with a 5\% margin of error. 
Subsequently, we manually identified VFCs within the sampled NVD records.
Three annotators independently identified VFCs from the NVD and meticulously documented their processes. 
For all categories of NVD records, they prioritized reviewing Git references and references tagged with \texttt{Patch}. 
If VFCs could not be located within these references, they proceeded to examine the remaining non-Git references and those without the \texttt{Patch} tag.
All three authors are research engineers with over five years of programming experience, ensuring a high level of expertise in the identification process.

\subsubsection{RQ2: Finding VFCs from Reference}
We directly extract commits from the Git references (Categories 1 and 2) as VFCs. 
There are also cases where Git references are issues or pull request links. 
For these cases, we extract all the commits inside these issues or pull request links by using GitHub, GitLab, or Bitbucket APIs and consider them as VFCs.

For NVD records without Git references (Category 3 and 4), we still explore the non-Git references by designing a VFC filtering heuristic that leverages domain-specific rules to refine the identification of true VFCs. 

To evaluate the precision of these automatically extracted candidate VFCs, we conducted a manual analysis to verify their correctness. 
For this purpose, we sampled a representative subset of NVD records, ensuring a 95\% confidence level with a 5\% margin of error.

\subsubsection{RQ3: Finding VFCs from External Security Databases}
We build an automated pipeline that begins with extracting CVE identifiers from NVD records. 
We then search these CVE identifiers across six common external security databases, including Snyk~\cite{snyk}, GitHub Advisory~\cite{advisories}, Ubuntu Security~\cite{ubuntu_db}, Nifi Apache Security~\cite{nifi_db}, Django Security~\cite{django_db}, and OSV Dev~\cite{osv}. 
Snyk is a well-established vulnerability database that provides detailed information on vulnerabilities across multiple programming languages and frameworks. 
GitHub Advisory is a platform that aggregates security advisories from various sources, offering a centralized repository of vulnerability information. 
Ubuntu Security, Nifi Apache Security, and Django Security are project-specific security resources that provide valuable insights into vulnerabilities affecting widely used open-source software. 
OSV Dev is a community-driven vulnerability database that collects and curates vulnerability information from multiple sources.

To assess the quality of the VFCs collected from the external security databases, we further conduct a manual assessment on a subset of VFCs with a 95\% confidence interval and 5\% margin of error.

\subsubsection{RQ4: Finding VFCs from GitHub Repositories}
To answer this RQ, we revisit the effectiveness of two state-of-the-art VFC search techniques, Prospector~\cite{prospector} and PatchFinder~\cite{li2024patchfinder}, in identifying VFCs for NVD records in their relevant GitHub repositories. 
Developed concurrently, these methods have not been directly compared until now. To assess their precision in VFC extraction, we first apply both techniques to our curated \texttt{After-March-2024} dataset. 
As Prospector outperforms PatchFinder on this dataset, we proceed to test Prospector further on the \texttt{Implicit VFCs} subset. 
Finally, we deploy Prospector across all implicit NVD records - those lacking patch-tagged Git references.

While alternative techniques like VCMatch~\cite{wang2022vcmatch}, PatchScout~\cite{patchscout}, and VFCFinder~\cite{dunlap2024vfcfinder} exist, we chose Prospector and PatchFinder for their alignment with our study's needs. PatchFinder has demonstrated superior performance over VCMatch and PatchScout in its evaluation~\cite{li2024patchfinder}. Although VFCFinder shows promise, it is less suitable for NVD records, as its inference phase relies on inputs formatted for GitHub GHSA or the OSV Vulnerability Database.

\subsubsection{RQ5: Comparison Among Different Sources}
In prior RQs, we have run different methods on NVD references, external security databases, and GitHub repositories, and we also conducted separate manual verification processes. In this RQ, we first compare the precision of each to draw comparisons.
Next, we compare the coverage of VFCs extracted from these sources.
We calculate the proportion of NVD records for which VFCs are found in each source and identify the unique contributions of each approach. 
Finally, we analyze the overlap between the VFCs extracted from the three sources to determine the extent to which they complement or duplicate each other. 
We use Venn diagrams to visualize and quantify the overlap, providing a clear picture of how these sources contribute to the overall set of VFCs.

\subsection{Dataset Collection}
To collect data for this study, we implemented a script to automate the retrieval of detailed information about NVD records using the NVD API up to March 28th, 2024.
This tool streamlines the identification of VFCs by extracting key details from the NVD. 
Specifically, in our study, we mainly leverage the following pieces of information: (1) descriptions, (2) references, and (3) Common Platform Enumeration (CPE) configurations.

The descriptions offer critical insights into the nature of each vulnerability, including the potentially affected code segments and libraries.
The references section includes links to technical records, vendor advisories, and, most importantly, patches or updates, which often contain or directly point to the relevant VFCs. 
Finally, the CPE configurations identify the specific affected products and versions.

\subsection{Manual VFCs Verification}
% In this section, we describe the details of our manual VFCs verification process that we perform in Section~\ref{sec:vfc_verify_rq2} and Section ~\ref{sec:vfc_verify_rq3}.
In this section, we describe the details of our manual VFCs verification process that we perform in Section~\ref{sec:vfc_verify_rq2} and Section ~\ref{sec:vfc_verify_rq3}, which is guided by clear Standard Operating Procedures (SOPs)~\cite{sop1, sop2} to ensure reproducibility, consistency, and objectivity. These SOPs include specific rules for manual annotation, a mechanism for conflict resolution among annotators, and steps taken to minimize bias, all of which are integrated into the verification workflow as follows.
For the manual annotation rules, annotators are required to examine multiple evidence sources for each candidate, including the NVD record (description, CWE, and affected versions/CPE when available), the complete code diff and commit message, and any linked artifacts (e.g., pull requests or issue-tracker tickets). They then verify temporal and version consistency by comparing the commit timestamp and version/tag information against the affected version ranges in NVD (if present), the project’s official release notes for fixed versions, and NVD CPE configuration data. Finally, they perform logical vulnerability matching to ensure the code changes plausibly address the vulnerability described in the NVD: the change should modify the relevant vulnerable component/location, apply a fix type consistent with the vulnerability class (e.g., input validation or sanitization for XSS), and precede (and be included in) the release that introduces the fixed version. A candidate is labeled as a true VFC only when these criteria are satisfied.
To handle conflict resolution among annotators, each candidate is annotated independently by multiple annotators; when labels disagree, the annotators revisit the evidence and discuss their rationales until reaching a single consensus label, which is then recorded with a brief justification.
To minimize bias and quantify reliability, our three annotators are research engineers with over five years of programming experience and follow the same SOP checklist for every candidate, and we further evaluate annotation reliability using Cohen’s kappa, observing strong inter-annotator agreement that supports the objectivity and robustness of the verification process.

\subsection{Evaluation Metrics}
In this section, we present the evaluation metrics we use to answer different RQs.

\subsubsection{Manual Identification}
In RQ1, to evaluate the effectiveness of manual annotators in identifying VFCs from NVD references, we define the success rate as follows:

\begin{align}
\text{Success Rate} = \frac{\# \text{NVD Records with at least 1 identified VFC}}{\# \text{Sampled NVD Records}} \times 100\%
\end{align}

Here, the numerator represents the number of NVD records where at least one VFC was successfully identified by the annotator, while the denominator corresponds to the total number of NVD records in the sampled dataset. 
This metric provides a quantitative measure of how successfully annotators can identify VFCs within the given sample.

\subsubsection{Automated Identification}
In RQ2-4, we have run different automated methods to extract VFCs.
To measure their effectiveness, we have performed manual verification on a sample of NVD records and the identified VFCs.

We report precision metrics from two perspectives:
\begin{itemize}
    \item $\text{SuccessRate}_{\text{Records}}$:  Accuracy of mapping NVD records to at least one true VFC.
    \item $\text{Precision}_{\text{VFCs}}$: Accuracy of identified VFC candidates being true VFCs.
\end{itemize}

\begin{align}
    \text{SuccessRate}_{\text{Records}}=\frac{\# \text{NVD Records with at least 1 confirmed true VFC}}{\# \text{Sampled NVD Records}} \times 100\%
\end{align}

\begin{align}
    \text{Precision}_{\text{VFCs}}=\frac{\# \text{Confirmed true VFCs}}{\# \text{Candidate VFCs}}\times 100\%
\end{align}

\subsubsection{GitHub Repository Search Method Comparison}
For RQ4, we use Recall@K (Formula~\ref{formula:recall}) to compare the performance of PatchFinder and Prospector on the \texttt{After-March-2024} dataset. 
Recall@K is also used by the authors of PatchFinder. 
It measures the proportion of true VFCs retrieved within the top-K predictions, with higher values indicating better performance.

\begin{align}
\label{formula:recall}
\text{Recall@k} = \frac{\# \text{True VFCs in Top-k predictions}}{\# \text{All VFCs}}
\end{align}

\subsubsection{Comparison Among Different Sources}
For RQ5, we compute coverage to show how many NVD records can be successfully mapped to at least one VFC by combining all the sources.

\begin{align}
\label{formula:coverage}
\text{Coverage} = \frac{\# \text{NVD Records with at least 1 identified VFC}}{\# \text{All NVD Records}}
\end{align}

\subsection{Implementation Details}

\subsubsection{RQ1: Manual Investigation}
At the beginning of the VFC identification process for each NVD record, three annotators, with at least five years of programming experience and five years of experience in security research, are instructed to read the NVD description and CWE information to gain a basic understanding of the vulnerability.  
Following this, they proceed to review the CPE sections to gather information about the affected projects. 
Participants then attempt to identify VFCs using all available resources, typically beginning with the references section of the NVD, which is the most probable source of VFCs.
When examining the references, participants are advised to prioritize them in the following order: patch-tagged Git links, non-patch-tagged Git links, patch-tagged non-Git links, and non-patch-tagged non-Git links. If no VFCs are found within the NVD references, participants are encouraged to extend their searches to other online sources. 
However, to ensure efficiency, the search for each CVE is limited to a maximum of 20 minutes.
At the end of each search, participants are required to complete an answer sheet, providing details such as the time spent on the search (in seconds), the links to any identified VFCs, and the methods used to find them.

\subsubsection{RQ2: Finding VFCs from Reference}
\label{sec:automated_pipeline}

% \begin{figure*}[h]
% \centering
% \includegraphics[width=\textwidth]{figure/4-1-workflow.png}
% \caption{Automated pipeline for VFC extraction from reference}
% \label{fig:vfc_matcher}
% \end{figure*}

% \st{The heuristic begins with a web scraper to extract and analyze all references within each NVD record. 
% It then identifies and retrieves the inner references associated with these records to construct the NVD reference tree. 
% This tree is built by recursively extending each reference through its own linked references.
% Notably, we start with the NVD-provided references as the first level and expand the tree by their references. 
% For example, if an NVD reference points to a security advisory or bug tracker, the heuristic further explores the references within those linked resources to identify additional candidate VFCs. 
% In other words, the reference tree is formed by recursively traversing the references and their linked references to expand its search space, increasing the likelihood of encountering a VFC.
% The depth parameter controls the extent of this recursive exploration, allowing for a balance between thoroughness and computational efficiency. 
% At each level of the tree, the heuristic extracts relevant information from the references, such as Git commits or pull requests, which are then filtered based on predefined criteria to identify candidate VFCs. This iterative process continues until a specified depth is reached or a VFC is found.
% However, due to hardware and runtime constraints, our experiments limit the depth of the tree to two levels.
% }

%MR
In this RQ, we attempt to find VFCs by recursively exploring the provided reference links – an approach that primarily targets the NVD records in Categories 3 and 4, which lack direct Git references. This process follows a chain of references from one webpage to another starting from a target NVD record, traversing what we refer to as an `NVD reference tree’. Specifically, we define the NVD reference tree of a target NVD record as a hierarchical structure formed by treating the NVD record as a root node. Next, all pages pointed to by reference links on the NVD record are considered as Depth 1 nodes. We can continue the process to identify Depth N+1 nodes from the hyperlinks found in the pages corresponding to Depth N nodes. In addition, we also have recorded what links have been checked on the NVD reference tree, so that any hyperlink that has already been visited at a previous depth is ignored and not explored again. This process is essentially similar to the breadth-first traversal. 

\begin{figure*}[h]
\centering
\includegraphics[width=1\textwidth]{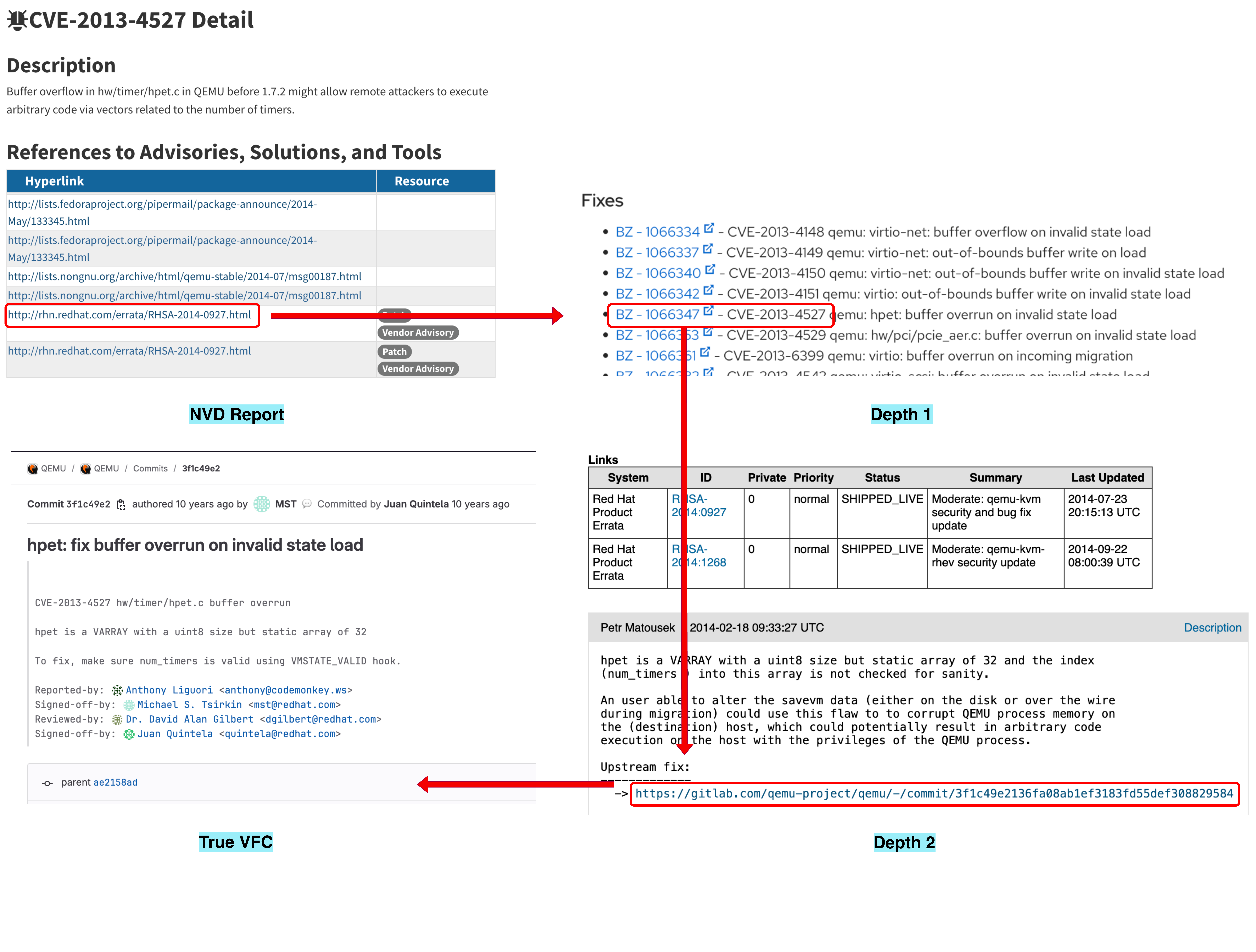} 
\caption{Concept of NVD reference tree  on the example of CVE-2013-4527}
\label{fig:nvd_ref_ex}
\end{figure*}

For instance, the figure~\ref{fig:nvd_ref_ex} illustrates this process for CVE-2013-4527: an initial NVD reference link points to a Red Hat Security Advisory (Depth 1). That advisory page, in turn, contains an "Upstream fix" link leading to the VFC from the specific QEMU project that resolves the vulnerability, which we find at Depth 2. This exploration allows us to uncover VFCs that are not immediately visible from the NVD's original page. 

To determine a practical depth limit for this recursive search, we performed an empirical study on a random sample of 100 NVD records from Categories 3 and 4, allowing the scraper to search with a virtually unlimited depth. Our analysis revealed that the vast majority of discoverable VFCs are found within the first two levels. Specifically, we found that 71.4\% of the true VFCs were identified at Depth 1 and 21.4\% were found at Depth 2. A negligible portion, only 7.2\% was discovered at a depth of 3 or greater, indicating sharply diminishing returns. Based on this data, we set the depth for our main automated pipeline to two levels. This decision represents an empirically-justified trade-off, ensuring our method captures the overwhelming majority (approximately 92.1\%) of findable VFCs while avoiding the significant computational cost and inefficiency of deeper, less promising searches.

% \st{We then apply a two-step filtering process once the references are extracted and organized. 
% First, we retain high-confidence references, such as commit and issue links, from Git-based repository hosting services like GitHub, GitLab, and Bitbucket. 
% Second, we cross-validate these references against the affected product information extracted from the CPE in the NVD record. 
% If a Git link is determined to originate from the same repository as the CPE, it is classified as a candidate VFC. 
% This validation step ensures that the identified commits are contextually relevant and directly associated with the reported vulnerability. 
% This process of cross-validation is applied to all the NVD records.}

To automate the VFC discovery from these references, we developed a two-step filtering heuristic. The design of this heuristic was directly informed by the strategies that proved most effective during our manual analysis, which is described in Section~\ref{sec:rq1_result}. First, we retain only high-confidence references, such as direct commit, issue, or pull request links from established Git-based services (e.g., GitHub, GitLab, Bitbucket). Our manual investigation revealed that these types of links have the highest probability of leading to a true VFC, whereas other links often lead to general advisories or product pages. Second, we cross-validate these references against the affected product information listed in the NVD record's Common Platform Enumeration (CPE). This step was crucial because our manual analysis uncovered instances where a reference link pointed to a commit in an incorrect or unrelated project. If a Git link's repository is successfully matched with the product information from the CPE, it is classified as a candidate VFC. This validation ensures the commit is contextually relevant to the reported vulnerability, significantly reducing false positives.

\subsubsection{RQ3: Finding VFCs from External Security Databases}
\label{sec:ex_source}
For resources with APIs, such as the GitHub Advisory Database and OSV Dev, we query their APIs for each CVE to extract Git links from their reference sections. 
For external security databases without APIs, we deploy a tailored web scraper for each, applying the filter heuristics outlined in Section~\ref{sec:automated_pipeline}. 
These heuristics confirm that collected VFCs originate from Git-based hosting services and align with the CPE information of the targeted NVD records. 
As with RQ1, we identified all potential candidate VFCs to evaluate their coverage and performed a manual analysis to verify their accuracy, thus assessing precision.

\subsubsection{RQ4: Finding VFCs from GitHub Repositories}
% For the rest of the NVD records that we cannot find their VFCs in the prior two steps, we further conduct a GitHub repository search.
% Particularly, we aim to answer the following RQ: 
% \begin{itemize}[left=0pt,label={}]
%     \item \textbf{RQ3:} \textit{How effective are existing VFC search techniques for identifying VFCs in GitHub repositories?}
% \end{itemize}

We provide a brief description and the specific implementation we conducted on the two tools, i.e., PatchFinder~\cite{li2024patchfinder} and Prospector~\cite{prospector} as follows.

\vspace{4px}
\noindent\textbf{PatchFinder~\cite{li2024patchfinder}:} \textit{(1) Description:} PatchFinder has two phases: Phase 1 (initial retrieval) and Phase 2 (re-ranking). In Phase 1, PatchFinder uses a hybrid patch retriever that combines lexical matching via TF-IDF and semantic matching using a pre-trained CodeReviewer~\cite{codereviewer} model. This phase narrows down the candidate set by extracting commits that are similar to the CVE descriptions in both lexical and semantic aspects. Phase 2 involves re-ranking these candidates using a fine-tuned CodeReviewer model to learn the semantic correlations between CVE descriptions and commits, to improve the accuracy of the patch identification.
\textit{(2) Implementation:} 
We first restructure our \texttt{After-March-2024} explicit dataset to have the same format as the authors' dataset.  
Specifically, for each NVD record found with VFCs, we follow PatchFinder's setting by randomly sampling 5,000 other non-patch commits from the same repositories to put into our dataset. 
We also construct another dataset containing NVD records before March 2024, to fine-tune Phase 2 of PatchFinder. 
The fine-tuning dataset includes all the explicit VFCs (patch-tagged Git references) associated with the 15,732 NVD records in Category 1.
After the fine-tuning, we run the inferences of all two phases of PatchFinder to get the output candidate VFCs, with the same settings mentioned in the PatchFinder paper.

\vspace{4px}
\noindent\textbf{Prospector~\cite{prospector}:} \textit{(1) Description:} Prospector, on the other hand, is a purely heuristic approach that is also developed to map vulnerability advisories, mainly NVD records, to the corresponding VFCs in public GitHub repositories. The methodology of Prospector involves retrieving vulnerability advisories, processing their descriptions to extract relevant keywords, file names, and method references, and then scanning source code repositories for commits that might correspond to these vulnerabilities. Using a set of heuristics, Prospector ranks candidate commits by evaluating the presence of certain indicators, such as the inclusion of vulnerability identifiers in commit messages or references to bug-tracking tickets and GitHub issues. This ranking is further refined by considering the changes made in the code itself. 

\textit{(2) Implementation:} 
We run the heuristic version of Prospector as described in their original paper~\cite{prospector}.
Although Prospector has a large language model (LLM) support feature in their GitHub repository~\cite{prospector:online} to enhance the quality of the results, we decided not to use it for two reasons. 
First, this feature is a wrapper for calling APIs of commercial LLMs, which is not cost-effective to run on a large database such as NVD. 
Secondly, to date, the authors of Prospector have not published any results related to this LLM-support feature, as the evaluation from their paper is conducted on the only heuristic version of Prospector. 
After getting the candidate VFCs output from Prospector, we only keep VFCs with Prospector scores over 60 to ensure the quality and reduce manual effort to check their validity. 
We also modified Prospector heuristic rules to not prioritize commits found in GitHub issues and release links, as we have manually observed many commits in these cases are not VFCs. 
To run Prospector for each NVD record, we need to provide the CVE ID and its corresponding GitHub repository link as inputs.

\vspace{4px}
\noindent\textbf{Evaluation Datasets}: It is important to note that existing works have primarily been evaluated on datasets of \textbf{known} VFCs, which were already identified as VFCs at the time of writing. 
However, the potential for data leakage remains unclear, particularly given that Prospector relies on heuristic rules; they might inadvertently evaluate data where those heuristic rules were initially formulated.

Our objective, in contrast, is to discover new and \textit{hidden} VFCs - those that were unlabeled at the time of their publication. 
To this end, we focus on assessing performance specifically on such VFCs. 
To achieve this, we construct two datasets:

\begin{itemize}[leftmargin=0pt, itemindent=*]
    \item \texttt{After-March-2024}: To evaluate the generalizability of PatchFinder and Prospector, we constructed a dataset comprising NVD records from Category 1—those featuring at least one Git reference tagged with \texttt{Patch}. This dataset includes records published after the release of these two tools in March 2024, extending through September 2024.
    We only include the NVD records belonging to Category 1, i.e., containing at least one patch-tagged Git reference; the rationale is that we achieve the highest precision in terms of NVD records and VFCs in both manual (RQ1) and automated identification (RQ2). This dataset aims to evaluate the tools automatically since we already have the ground-truth VFCs.
    \item \texttt{Implicit VFCs}: The second dataset comprises NVD records from Categories 2–4, which lack Git references tagged as \texttt{Patch}. Although these categories encompass 219,609 NVD records in total, Prospector requires two input parameters for each record: the CVE ID and the GitHub URL of the corresponding source repository~\cite{prospector}. Consequently, we include only those implicit NVD records for which public GitHub repository links are available, resulting in a dataset of 99,345 records, termed the \texttt{Implicit VFCs} dataset. The evaluation focuses exclusively on these implicit records from Categories 2, 3, and 4, as verifying the correctness of candidate VFCs for these records is inherently more complex than for the previous dataset, which benefits from established ground-truth VFCs.
    Since these NVD records lack explicit ties to ground-truth VFCs, we rely on a manual verification process. 
    We randomly selected 300 NVD records for which Prospector has generated VFC predictions and documented the outcomes of manual verification.
    Annotators review the candidate VFCs generated by the method under study to assess their validity.
\end{itemize}

\section{Results}
\label{sec:results}
This section presents the results of our empirical study, addressing each RQ in sequence to evaluate the feasibility and challenges of mapping NVD records to their VFCs.

\subsection{RQ1: Manual Investigation}
\label{sec:rq1_result}
\begin{table*}[h]
\centering
\caption{Statistics of NVD record categories and manual VFC extraction results.}
\begin{tabular}{lrrrrr}
\toprule
\textbf{Category}                                           & \textbf{\# Records} & \textbf{\# Sampled} & \textbf{A\#1} & \textbf{A\#2} & \textbf{A\#3}         \\ \midrule
C1: Git Ref. with \texttt{Patch} tag       & 15,732              & 375                 & 369 (98.4\%)  & 369 (98.4\%)  & 369 (98.4\%) \\
C2: Git Ref. without \texttt{Patch} tag    & 2,813               & 339                 & 294 (86.7\%)  & 302 (89.1\%)  & 299 (88.2\%) \\
C3: No Git Ref. but has \texttt{Patch} tag & 51,958              & 382                 & 65 (17\%)     & 55 (14.4\%)   & 59 (15.4\%)  \\
C4: No Git Ref. and no \texttt{Patch} tag  & 164,838             & 384                 & 38 (9.9\%)    & 32 (8.3\%)    & 35 (9.1\%)   \\ \bottomrule
\end{tabular}%
\label{tab:manual}
\begin{tablenotes}
\small
\item \textit{Note:} The ``Sampled'' column represents the NVD records selected for manual identification. ``A\#1'', ``A\#2'' and ``A\#3'' indicate the number of NVD records where Annotator 1, Annotator 2 and Annotator 3 successfully identified VFCs, respectively. Values in parentheses denote the success rate. ``Ref.'' refers to references.
\end{tablenotes}
\end{table*}

Table~\ref{tab:manual} shows the results of manual VFC identification across the four NVD record categories defined in Section~\ref{sec:nvd_categories}.
In total, we have crawled 235,341 NVD records.
The table also includes the results of the manual investigation.

\subsubsection{NVD Records with Git References}

\begin{itemize}[leftmargin=0pt, itemindent=*]
\item \textbf{Category 1: Git references with \texttt{Patch} tag.} All three annotators successfully identified VFCs in 98.4\% of the 375 sampled records, reflecting high reliability.
A small proportion of failed cases are the broken and inaccessible Git references, which are potentially due to repository deletion, renaming, privatization, or commit history changes (e.g., rebasing or squashing).
\item \textbf{Category 2: Git references without \texttt{Patch} tag.}
Annotators achieved success rates of 86.7\% (294/339), 89.1\% (302/339) and 88.2\% (299/339), with a Cohen's kappa of 0.82 indicating strong inter-annotator agreement.
Despite the absence of \texttt{Patch} tags, most Git references proved to be true VFCs upon verification.
The few failed cases of this category are due to irrelevant Git references, such as links to issues, releases, or repository home pages, rather than actual VFCs.
This may be explained by the fact that these Git references were submitted to the NVD, but the NVD has not yet analyzed them, resulting in the absence of \texttt{Patch} tags. 
This also suggests that future works can leverage these references as VFCs with high confidence.
\end{itemize}

\find{
\textbf{Insight 1:} In NVD records with Git references, the presence of the \texttt{Patch} tag is not critical for identifying VFCs. Commits from both tagged and untagged Git references serve as reliable candidate VFCs.
}

\subsubsection{NVD Records without Git Reference}

\begin{itemize}[leftmargin=0pt, itemindent=*]
\item \textbf{Category 3: No Git references but has \texttt{Patch} tag.} 
For NVD records with patch-tagged non-Git references, the three annotators encountered greater difficulty, with success rates dropping to 17\%, 14.4\% and 15.4\%, and a Cohen's kappa of 0.75, indicating moderate agreement. 
This highlights the increased effort required to identify VFCs in this category, even if they were confirmed to be highly relevant to VFCs. 
\item \textbf{Category 4: No Git references and no \texttt{Patch} tag.} Similarly, for NVD records with non-patch-tagged non-Git references, three annotators identified VFCs at even lower rates of 9.9\%, 8.3\% and 9.1\%, with a Cohen's kappa of 0.78. 
Merged together, the annotators successfully identified VFCs in only 41 out of the 384 NVD records (10.7\%), reflecting considerable variability in annotator agreement and the significant challenges of identifying VFCs without Git references or \texttt{Patch} tags.
\end{itemize}

The main reason for the lower number of VFCs found in Categories 3 and 4 compared to Categories 1 and 2 is that, without Git references, the annotators had to rely on more indirect and time-consuming methods to find VFCs. To be more specific, finding VFCs in Categories 3 and 4 required the annotators to go through different non-Git references to get every hint that could lead to VFCs.
For example, there is a case where one of the annotators found the release tag of the fixed version for an NVD record. He then used this release tag to trace back the latest commit before the tag and manually double-checked again to ensure this commit was the true VFC. Another reason is that a significant portion of the affected products associated with these NVD records are not open-source, making it impossible to trace corresponding VFCs.

\find{
\textbf{Insight 2:} For NVD records without Git references, the \texttt{Patch} tag does not significantly aid VFC identification.
The challenges in identifying VFCs from mining NVD records with non-Git references lie in the limited availability of publicly accessible version-control data, the complexity of cross-referencing multiple documents, and the reliance on extensive manual investigation. 
Nonetheless, the commits extracted from these records can still be considered potential VFCs.
}

\subsubsection{Potential Sources of VFCs}
One of the main goals of the manual analysis is to identify sources of VFCs beyond patched Git references in the NVD. 
This approach helps address gaps and increases the number of VFCs that can be identified in the NVD, particularly in the absence of Git references. 
To achieve this, the annotators further extract information from the NVD records belonging to either Category 3 (Patch-tagged non-Git reference) or Category 4 (Non-Patch-tagged non-Git reference).
The annotators then perform open card sorting~\cite{spencer2009open_card} to categorize the sources used to identify VFCs. 
At the end of this process, three categories were established: \textit{By references}, \textit{By Git}, and \textit{By external resources}. 
In the following section, we will first describe each category with examples and then present their distributions.

\begin{figure*}[h]
\centering
\includegraphics[width=1\textwidth]{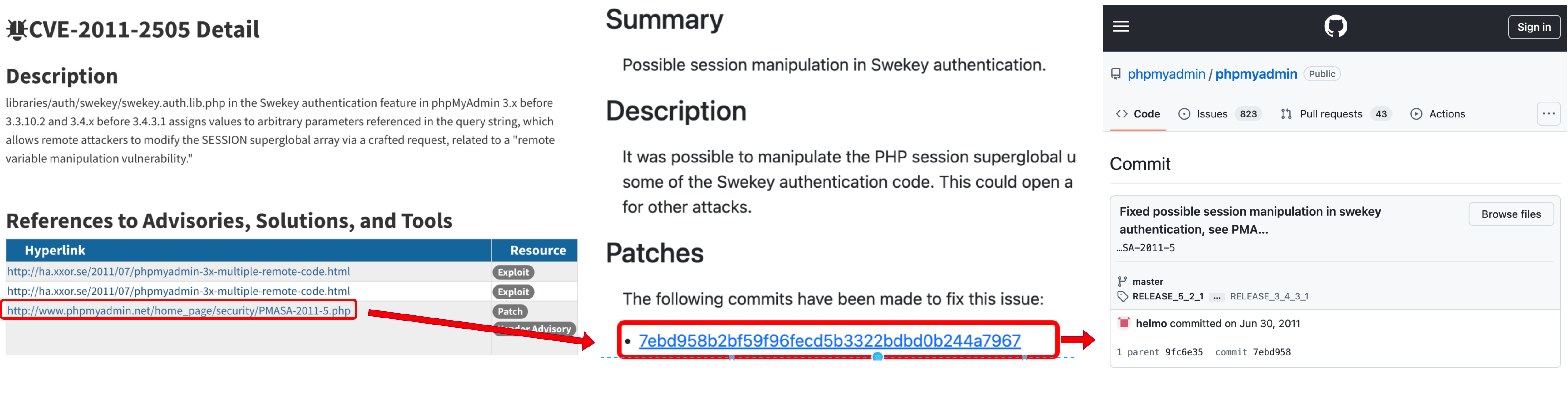} % Adjust width to 80% of text width
\caption{Example of manually finding the VFC for CVE-2011-2505 by references}
\label{fig:example_ref}
\end{figure*}

\textbf{By references} refers to the process in which the annotators review all references from the NVD, such as URLs, advisories, or third-party records, to identify VFC links. 
The annotators may also follow additional links from these references and repeat the search process. However, due to the time constraint of 20 minutes per CVE, annotators are limited to searching a maximum of three inner links from the original references. An example of this process is illustrated in Figure \ref{fig:example_ref}: In the reference section of NVD for CVE-2011-2505, we examine all the reference URLs and identify a GitHub commit link from one of the non-Git references.
Upon manual review, we confirm that this commit from the public repository of the affected product (phpMyAdmin) addresses the vulnerability described in the CVE (Swekey-related session manipulation). 
Therefore, we can confirm that this commit is the VFC for CVE-2011-2505.

\begin{figure*}[h]
\centering
\includegraphics[width=1\textwidth]{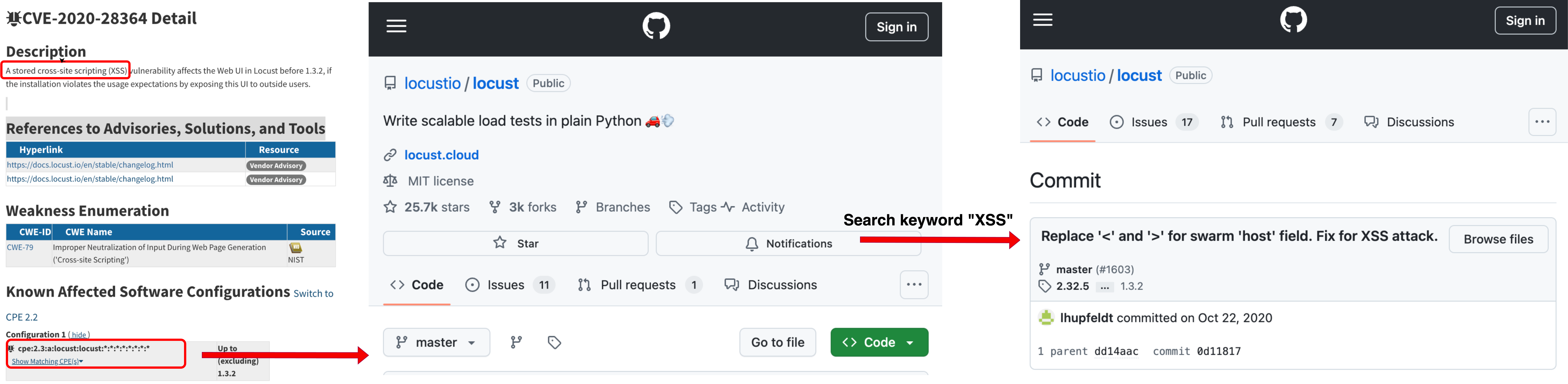} 
\caption{Example of manually finding the VFC for CVE-2020-28364 by Git}
\label{fig:example_git}
\end{figure*}

\textbf{By Git} refers to the process where, for each CVE, the annotators extract key keywords from its NVD description and CWE description. The annotators then attempt to explore the public Git repository link for the affected product and search its Git commits for messages containing one or more of the extracted keywords. From this set of Git commits, the annotators conduct a final manual review to identify the VFCs. Figure \ref{fig:example_git} illustrates an example of manually identifying the VFC for CVE-2020-28364 through Git repositories. In this example, as there is no candidate VFC that could be found in the references, we try to find if there are any accessible public Git repositories from the affected products of this CVE.
Fortunately, through the CPE information of NVD, we can find public GitHub repositories of Locust, which is the affected product. 
Since the vulnerability of the CVE is cross-site scripting, we search for a set of keywords related to cross-site scripting in this repository and find a commit whose message contains ``XSS''. After that, we again conduct a manual check to verify the correctness of this commit. To be more detailed, we see that the code diff of this commit fixing the XSS vulnerability in the ``web.py'' file particularly fits the CVE description, which describes an XSS vulnerability in the Web UI of Locust. 
In addition, the author date of this commit, which is October 22nd, 2020, is also very close to the CVE published date, which is November 9th, 2020. 
With these justifications, it can be confirmed that this commit is indeed the true VFC of CVE-2020-28364.

\textbf{By external resources} refers to the process in which, for each CVE, the annotators extend their search beyond the NVD by exploring other resources, websites, and security advisories, such as Bugzilla~\cite{bugzilla} or the Open Source Vulnerability Database (OSV)~\cite{osv}, among others. 
These external resources may contain VFCs or crucial information leading to VFCs that are not covered by NVD.

\begin{figure*}[h]
\centering
\includegraphics[width=1\textwidth]{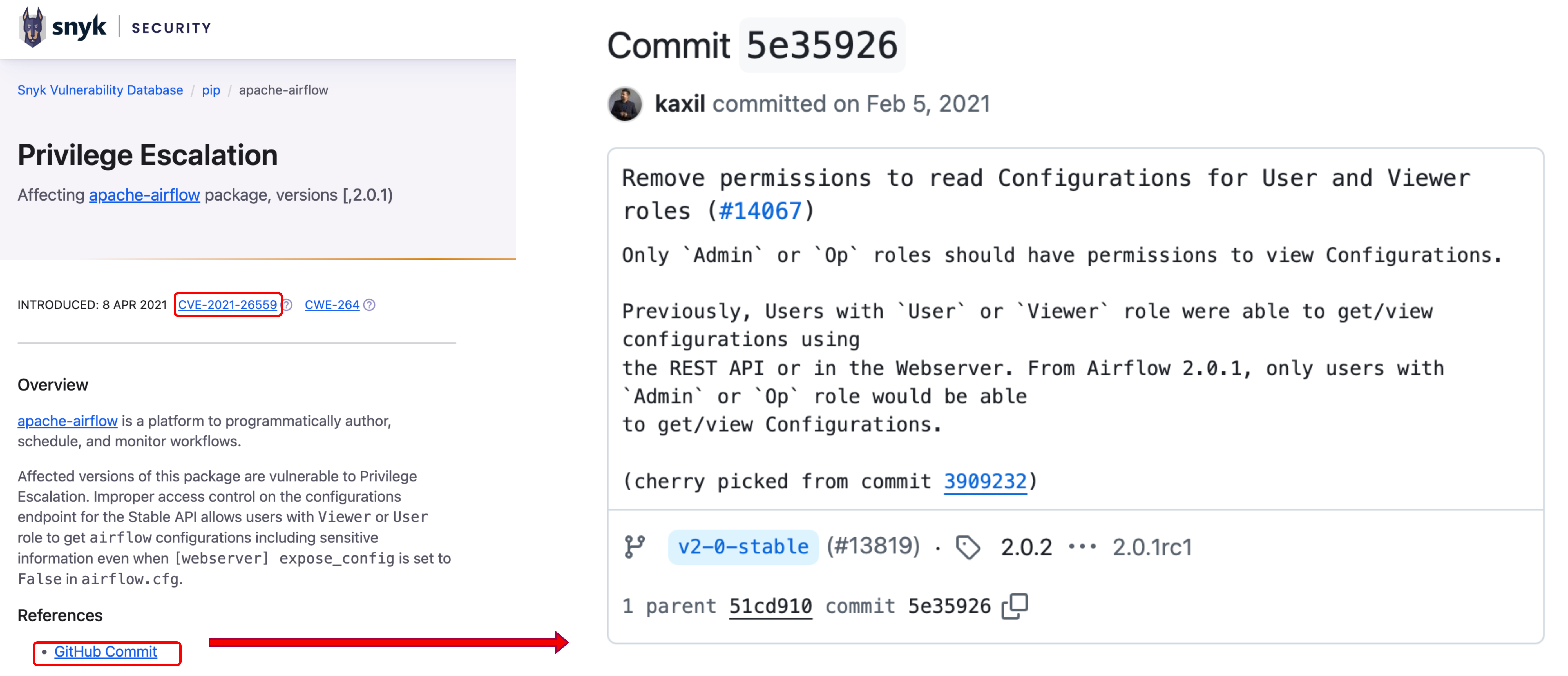} 
\caption{Example of manually finding the VFC for CVE-2020-28364 through Snyk.io, an external advisories}
\label{fig:example_ex}
\end{figure*}

An example of using this approach for CVE-2020-28364 is shown in Figure \ref{fig:example_ex}. 
In this example, we cannot find any candidate VFC from the two above approaches. Therefore, we shift to external advisories beyond NVD, starting with Snyk.io~\cite{snyk}.
In the reference section of Snyk advisories for this CVE, we find a GitHub commit link. Through the manual check, it is confirmed that this commit is indeed the VFC for CVE-2020-28364. 

\begin{figure*}[t]
\centering
\includegraphics[width=0.8\textwidth]{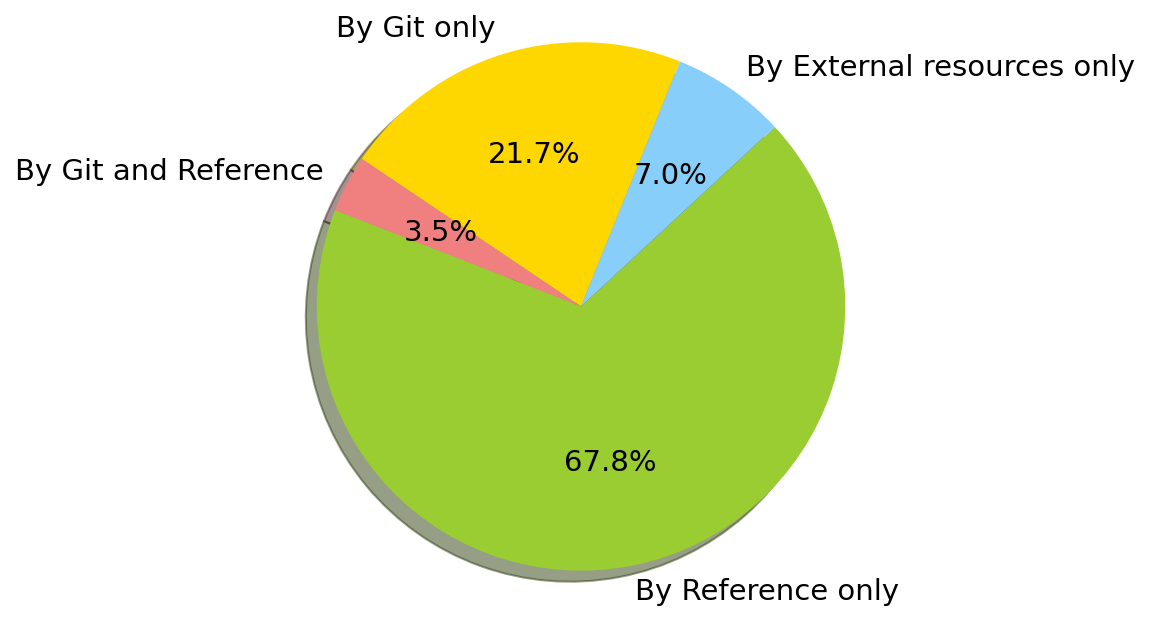}
\caption{Combined statistic on possible sources by three annotators}
%\rw{need to remove Example Pie Chart in figure}}
\label{fig:pie_source}
\end{figure*}

Figure \ref{fig:pie_source} shows the distribution of VFC sources used by three annotators to identify VFCs. 
The data shows that the majority of NVD records (67.8\%), have VFCs identified exclusively through NVD references, highlighting the significant role NVD references play in locating VFCs.
VFCs found solely in Git rank second, accounting for 21.7\% NVD records, indicating that while Git sources are essential, they are less frequently the sole resource for VFC discovery compared to NVD references. 
Furthermore, annotators find VFCs for only 7.0\% NVD records exclusively through external resources. 
This suggests that while external resources may be useful for identifying VFCs, their sparsity makes it difficult for annotators to consistently identify the correct VFCs, resulting in their lower significance.

On the other hand, the statistics also reveal that a smaller subset of NVD records have VFCs identified from multiple sources. 
This refers to cases where one annotator identifies VFCs for an NVD record from one source. 
In contrast, the other annotator identifies VFCs for the same NVD record from a different source. 
Specifically, 3.5\% NVD records have VFCs identified from both Git and NVD references. 
This suggests that identifying VFCs from multiple sources may increase the likelihood of discovering true VFCs.

\find{
\textbf{Insight 3:} Manual analysis identified three VFC source types: \textit{By references, By Git, and By external sources}. NVD references dominated (67.8\%), followed by Git  (21.7\%), with external resources contributing minimally (7.0\%).
}

\subsubsection{Failure causes}
In this manual study, we also encountered several challenges that hindered the successful identification of VFCs for certain NVD records. 
First, during the manual investigation for RQ1, the annotators had documented the specific reasons for each failure on their answer sheets. Next, using an open card sorting ~\cite{spencer2009open_card} methodology, the annotators grouped the documented failure reasons based on thematic similarity. They then collaboratively assigned descriptive labels to each category. Open card sorting is a qualitative analysis technique where researchers group individual data points to identify emergent thematic structures, and it is a well-established method for categorizing data followed by many prior studies~\cite{ wan2017bug, widyasari2025explaining, zheng2025towards}. Through this process, we identified three primary categories: repository accessibility, documentation and reference issues, and commit-related factors.

Firstly, we were unable to find the repositories of affected products in some cases, particularly for enterprise products or those not using Git as their version control system. Secondly, the information provided in NVD descriptions and references was sometimes insufficient, too general, or redundant, making it difficult to locate the corresponding VFCs effectively. Lastly, commit-related issues posed significant obstacles, such as the possibility of VFCs being squashed or rebased, rendering them unavailable for analysis. 
Additionally, some commit messages were too vague or not directly related to the vulnerability. 
In contrast, in other cases, vulnerability fixes were part of multi-purpose commits, further complicating the identification process. 
These findings not only highlight the limitations of solely relying on NVD references for VFC identification but also provide valuable insights into the characteristics and challenges associated with the NVD-VFC mapping problem. By understanding the factors that influence the ease or difficulty of identifying VFCs for different types of NVD records, future research can develop more targeted and effective approaches to address these challenges.

\find{
\textbf{Insight 4:} The manual investigation faced challenges in identifying VFCs for certain NVD records, with varying difficulty based on project type and version control system. The main obstacles were repository accessibility, insufficient documentation and references, and commit-related issues
}

\subsection{RQ2: Finding VFCs from NVD References}
\begin{table*}[h]
\centering
\caption{Statistics of VFCs extracted from NVD references using our filter heuristics.}
\label{table:4_1_data}
\begin{tabular}{lrrrrr}
\toprule
\multirow{2}{*}{\textbf{Category}}  & \multirow{2}{*}{\textbf{\# Webpages}} & \multirow{2}{*}{\textbf{\# Records}} & \textbf{\# Identified} & \textbf{\# Candidate} \\
&   &   & \textbf{Records} & \textbf{VFCs}     \\ 
\midrule
C1: Git Ref. with \texttt{Patch} tag  & \_  & 15,732  & 15,732 & 23,377 \\
C2: Git Ref. without \texttt{Patch} tag & \_  & 2,813 & 2,813 & 5,872 \\
C3: No Git Ref. but has \texttt{Patch} tag & 61,238 & 51,958 & 833 & 1,303 \\
C4: No Git Ref. and no \texttt{Patch} tag  & 314,550 & 164,838 & 982 & 1,390 \\
\midrule
Total   & 375,788  & 235,341   & 20,360  & 31,942 \\ 
\bottomrule
\end{tabular}
\begin{tablenotes}
\small
\item \textit{Note:}  ``\# Webpages'' and ``\# Records'' denote the number of webpages and NVD records crawled. ``\# Identified Records'' denotes the number of NVD records associated with candidate VFCs.
\end{tablenotes}
\end{table*}

\subsubsection{Automated Extraction Result.}
Table \ref{table:4_1_data} presents the numerical statistics of VFCs extracted by our automated pipeline across various categories of NVD records. 
The proportion of NVD records with at least one extracted VFC was calculated relative to the total number of processed records. 
In total, the pipeline was applied to 235,341 NVD records.

For NVD records containing at least one Git reference (Categories 1 and 2), commits were extracted from those references. Specifically, for patch-tagged Git references (Category 1), the pipeline identified 23,377 candidate VFCs from 15,732 NVD records. For non-patch-tagged Git references (Category 2), it extracted 5,872 candidate VFCs from 2,813 NVD records.
For NVD records containing only non-Git references (Categories 3 and 4), the pipeline successfully crawled 61,238 and 314,550 web pages corresponding to 51,958 and 164,838 NVD records, respectively. From these, it identified 1,303 and 1,390 candidate VFCs, covering 833 and 982 NVD records, respectively.
In total, the automated pipeline extracted 31,942 VFC candidates from 20,360 NVD records.

\begin{table*}[h]
\centering
\caption{Manual verification result: VFCs extracted from references across different categories of NVD records.}
\label{tab:4_1_result}
\begin{tabular}{lrrrr}
\toprule
\textbf{Category} & \textbf{\# Records} & \textbf{\# True Records} & \textbf{\# VFCs} & \textbf{\# True VFCs} \\ 
\midrule
C1: Git Ref. with \texttt{Patch} tag & 375 & 369 (98.4\%)  & 402 & 391 (97.3\%) \\
C2: Git Ref. without \texttt{Patch} tag  & 316 & 283 (89.6\%) & 361 & 303 (83.9\%) \\
C3: No Git Ref. but has \texttt{Patch} tag & 264  & 225 (85.2\%) & 351 & 284 (80.9\%) \\
C4: No Git Ref. and no \texttt{Patch} tag  & 277  & 224 (80.9\%) & 301 & 253 (84.1\%) \\ 
\midrule
Total  & 1232  & 1101 (89.4\%) & 1415  & 1231 (87.0\%) \\ 
\bottomrule
\end{tabular}%
\begin{tablenotes}
\small
\item \textit{Note:} ``\# Records'' indicate the number of sampled records; ``\# True Records'' refers to the number of records manually confirmed to contain at least one true VFC, with the number in parentheses representing the $\text{SuccessRate}_\text{Records}$. ``\# VFCs'' represents the total number of candidate VFCs; ``\# True VFCs'' represents the subset of those confirmed as true VFCs, with the number in parentheses representing the $\text{Precision}_\text{VFCs}$.
\end{tablenotes}
\end{table*}

\subsubsection{Manual Verification Result.}
\label{sec:vfc_verify_rq2}
Table~\ref{tab:4_1_result} presents the manual evaluation results.
In detail, Git references with \texttt{Patch} tags (Category 1) demonstrate the highest precision, with 98.4\% of records containing at least one true VFC and 97.3\% of candidate VFCs being verified as true. 
This underscores the reliability of Git references tagged with \texttt{Patch} for VFC identification.
For Git references without \texttt{Patch} tags (Category 2), 83.9\% of candidate VFCs are confirmed as true VFCs, resulting in 89.6\% of records containing at least one true VFC. This indicates that while Git references without \texttt{Patch} tags are still highly useful, they require more manual effort for validation compared to their patch-tagged counterparts.

In the non-Git category (Category 3), the results were also strong, with 284 true VFCs (80.9\%) identified out of a total of 351 VFCs examined. This corresponds to 225 NVD records, representing 85.2\% of the total sample NVD records.
For the non-patch-tagged non-Git category (Category 4), a manual review of 301 VFCs revealed that 253 VFCs (84.1\%) were correctly identified as true VFCs, corresponding to 224 NVD records with at least one confirmed VFC (80.9\% of the total sample NVD records).

% Overall, the precision rates exceeding 80\% across all CVE categories demonstrate the effectiveness of our filtering heuristics in accurately identifying VFCs from NVD records.

In total, by merging the VFCs extracted from NVD references from all NVD record categories, we obtain a total of 31,942 VFCs across 20,360 NVD records, with an overall precision of 87.0\% and 89.4\% for VFCs and NVD records, respectively. 
Prior studies usually only focus on Git references (Categories 1 and 2), by adding extraction on non-Git references (Categories 3 and 4), we increase the size by 9.2\% in terms of VFCs and 9.8\% in terms of NVD records. 
These results demonstrate the potential of further extraction of the non-Git reference in improving vulnerability datasets. 
However, it is also worth noting that even combining all the NVD records that we found VFCs from NVD references, they still only account for 8.7\% of all NVD records, highlighting the need for further exploration and improvements.

\find{
\textbf{Answer to RQ2:} 
% Expanding beyond prior studies that focused solely on Git references (Categories 1 and 2), our inclusion of non-Git references (Categories 3 and 4) increases the dataset size by 9.2\% in VFCs and 9.79\% in NVD records. 
% Despite these improvements, the dataset represents only 8.66\% of all NVD records, underscoring the need for further research and refinement.
The automated extraction of VFCs from NVD references identifies a total of 31,942 VFCs from 20,360 NVD records, achieving the precision of 87.0\% and 89.4\%, respectively. 
However, the dataset represents only 8.7\% of all NVD records, underscoring the need for further research and refinement.
}

\subsection{RQ3: Finding VFCs from External Security Databases}
\begin{table}[h]
\caption{Statistics of NVD records and candidate VFCs extracted from external databases.}
\label{tab:4_2_data}
\begin{tabular}{lrr}
\toprule
\textbf{External Database}  & \textbf{\# NVD Records} & \textbf{\# Candidate VFCs} \\ \midrule
Snyk~\cite{snyk} & 18,521  & 27,583 \\
GitHub Advisory~\cite{advisories}   & 12,675 & 19,215      \\
Ubuntu Security~\cite{ubuntu_db}   & 3,167   & 4,022   \\
Nifi Apache Security~\cite{nifi_db} & 68  & 68  \\
Django Security~\cite{django_db}  & 142  & 301 \\
OSV Dev~\cite{osv}  & 697 & 855 \\ 
\midrule
Merged  & 18,985  & 29,254   \\ 
\bottomrule
\end{tabular}%
\end{table}

\subsubsection{Automated Extraction Result.}
Table \ref{tab:4_2_data} presents the data statistics of external databases in identifying additional candidate VFCs for NVD records. 
Among these sources, Snyk~\cite{snyk} and GitHub Advisory~\cite{advisories} contribute the most significant number of NVD records, with 18,521 and 12,675, respectively, resulting in 27,583 and 19,215 candidate VFCs. 
Other sources, such as Ubuntu Security~\cite{ubuntu_db}, Django Security~\cite{django_db}, and OSV Dev~\cite{osv}, provide smaller numbers of NVD records but still add valuable diversity to the dataset, offering additional candidate VFCs that the primary databases may not capture.   
To compare with other approaches, we merge the VFCs extracted from all the external sources together into a single dataset.
When different external sources provide different VFCs for the same NVD record, the union of these VFCs is included in the merged dataset. 
Overall, the combined dataset for external security databases covers 18,985 unique NVD records, resulting in a total of 29,254 candidate VFCs, highlighting the value of leveraging diverse external security databases to enrich the dataset.

\begin{table}[h]
\caption{Manual verification result: VFCs identified from external databases.}
\label{tab:4_2_result}
\begin{tabular}{lrrrr}
\toprule
\textbf{External Database} & \textbf{\# Records} & \textbf{\# True Records}  & \textbf{\# Candidate VFCs} & \textbf{\# True VFCs}  \\ \midrule
Snyk~\cite{snyk}   & 377    & 344 (91.2\%)  & 442  & 388 (87.7\%) \\
GitHub Advisory~\cite{advisories} & 373 & 352 (94.4\%) & 440 & 407 (92.5\%) \\
Ubuntu Security~\cite{ubuntu_db}  & 343 & 295 (86.0\%)  & 401 & 344 (85.8\%)  \\
Nifi Apache Security~\cite{nifi_db}  & 58  & 52 (89.7\%) & 58 & 52 (89.7\%) \\
Django Security~\cite{django_db} & 104 & 91 (87.5\%)  & 275 & 240 (87.2\%) \\
OSV Dev~\cite{osv} & 248  & 220 (88.7\%) & 296 &  256 (86.4\%) \\ 
\midrule
Merged &   1,472   & 1,334 (90.6\%) & 1,890 &  1,671 (88.4\%) \\ 
\bottomrule
\end{tabular}
\begin{tablenotes}
\small
\item \textit{Note:} ``\# Records'' indicate the number of sampled records; ``\# True Records'' refers to the number of records manually confirmed to contain at least one true VFC, with the number in parentheses representing the $\text{SuccessRate}_\text{Records}$. ``\# VFCs'' represents the total number of candidate VFCs; ``\# True VFCs'' represents the subset of those confirmed as true VFCs, with the number in parentheses representing the $\text{Precision}_\text{VFCs}$.
\end{tablenotes}
\end{table}

\subsubsection{Manual Verification Result.} 
\label{sec:vfc_verify_rq3}
Table \ref{tab:4_2_result} presents our manual assessment result.
For the Snyk.io Database, 442 VFCs are reviewed, 87.7\% confirmed as true VFCs, associating with 344 NVD records (91.2\% of the total sample NVD records). 
The GitHub Advisory Database shows similar precision, with 94.4\% (352 of 373) of NVD records containing at least one true VFC and 92.5\% (407 of 440) of individual VFCs verified as correct. 
Other sources, including Ubuntu, Apache NiFi, Django, and OSV Dev, also demonstrate high precision, with the percentage of true NVD records ranging from 86.0\% to 94.4\% and true VFCs from 86.4\% to 92.5\%.
These results highlight the high quality of VFCs identified from external security databases, particularly the GitHub Advisory Database and Snyk.io, which serve as valuable complements to the NVD in identifying implicit VFCs for NVD records.

% \begin{figure}[h]
%     \centering
%     \includegraphics[width=0.5\linewidth]{figure/venn_rq2_cve.png}
%     \caption{Venn diagram on NVD records with VFCs identified by NVD references and external security databases}
%     \label{fig:venn_rq2_cve}
% \end{figure}

% \begin{figure}[h]
%     \centering
%     \includegraphics[width=0.5\linewidth]{figure/venn_rq2_vfc.png}
%     \caption{Venn diagram on VFCs identified by NVD references and external security databases}
%     \label{fig:venn_rq2_vfc}
% \end{figure}

\subsubsection{Comparing with NVD}

\begin{figure}[ht]
    \centering
    % First subfigure
    \begin{subfigure}[t]{0.49\textwidth}
        \centering
        % Use width=\textwidth so the image fills the subfigure
        \includegraphics[width=\linewidth]{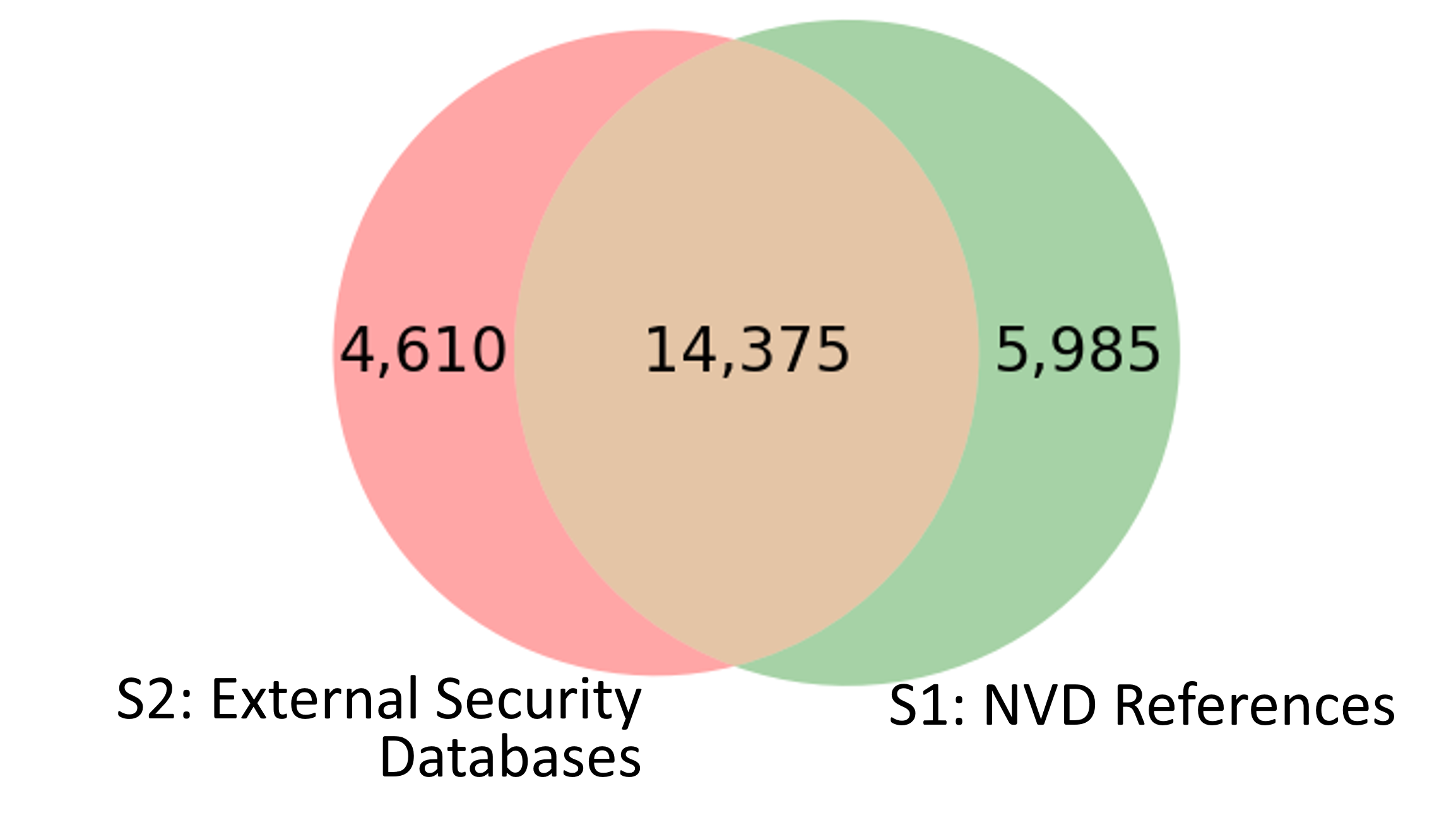}
        \caption{}
        \label{fig:venn_rq2_cve}
    \end{subfigure}
    \hfill
    % Second subfigure
    \begin{subfigure}[t]{0.49\textwidth}
        \centering
        \includegraphics[width=\linewidth]{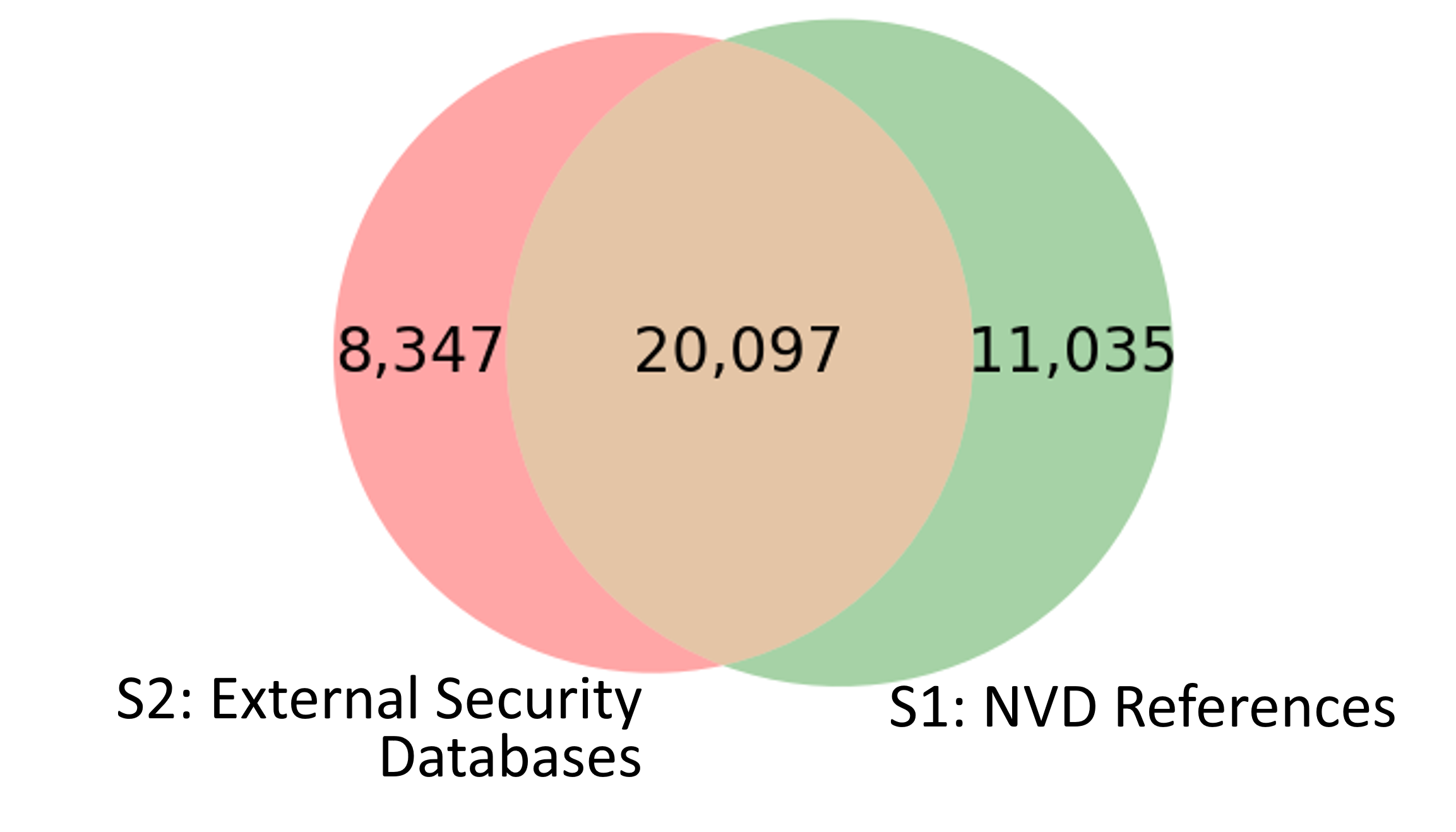}
        \caption{}
        \label{fig:venn_rq2_vfc}
    \end{subfigure}
    
    \caption{Venn diagrams comparing NVD records (a) and VFCs (b) between NVD References and External Security Databases.}
    \label{fig:venn_rq2}
\end{figure}

\label{sec:comparing_NVD_external}
Since the NVD and other external security databases exist independently, there may be questions about the overlap rate between the NVD records and VFCs across these different databases.
We, therefore, further compared the VFCs extracted from the NVD references and those extracted from other security databases.
We collectively refer to other security databases altogether as \textit{External Security Databases}.

Figures \ref{fig:venn_rq2_cve} and \ref{fig:venn_rq2_vfc} illustrate the overlap between VFCs identified through NVD references (green) and the combined set of VFCs extracted from all external security databases (red). 
In Figure \ref{fig:venn_rq2_cve}, 14,375 NVD records are shared between the two sources, while 4,610 and 5,985 NVD records are uniquely identified by external security databases and NVD references, respectively. 
Similarly, in Figure \ref{fig:venn_rq2_vfc}, 20,907 VFCs are shared, with 8,347 and 11,035 uniquely identified by external security databases and NVD references.  

Interestingly, this finding contrasts with our results from the manual analysis (Section~\ref{sec:rq1_result}), where the annotators were able to identify only 7.0\% number of VFCs from this approach, which is much fewer than 67.8\% number of VFCs identified through NVD references. 
We suspect that these differences arise from two factors. 

First, in the manual approach, annotators typically begin by examining NVD references before consulting external security databases. 
As a result, when a VFC is found in both sources, the annotators often attribute its discovery to NVD references, since it was identified there first. 
Second, while our automated approach is designed to replicate the manual search process, it is inherently limited in capturing the full complexity and flexibility of manual investigation within NVD references.
The automated reference scraping cannot explore all the potential connections in NVD references as comprehensively as human annotators, leading to fewer VFCs being found through NVD references in the automated approach. This limitation raises a new challenge in improving our automated VFC identification to further utilize the potential of NVD references for this task. 

\find{
\textbf{Answer to RQ3:} 
The automated extraction of VFCs from external security databases identifies a total of 29,254 candidate VFCs for 18,985 NVD records, achieving an average $\text{SuccessRate}_{\text{Records}}$ of 90.6\% and $\text{Precision}_{\text{VFCs}}$ of 88.4\%, respectively.
}

% \begin{table}[h]
% \centering
% \caption{Statistics of our dataset by combining Steps 1 and 2. }
% \label{tab:combined_dataset}
% \begin{tabular}{@{}lrrrr@{}}
% \toprule
% \textbf{Dataset} & \textbf{\# Records} & \textbf{\# VFCs} & \textbf{$\text{SuccessRate}_{\text{Records}}$} & \textbf{$\text{Precision}_{\text{VFCs}}$}\\ 
% \midrule
% S1: From NVD Reference & 20,360       & 31,942     & 89.4\% & 87.0\% \\
% S2: From External Databases   & 5,057    &  8,734  & 90.6\%  & 88.4\% \\ 
% \midrule 
% Full Dataset & 25,417  &   40,676   & 90.0\%       & 87.9\%  \\ 
% \bottomrule
% \end{tabular}
% \end{table}

\subsection{RQ4: Finding VFCs from GitHub Repositories}
\label{rq4_result}

\begin{table*}[h]
\centering
\caption{Recall@k results of Prospector and PatchFinder on \texttt{After-March-2024} dataset.}
\begin{tabular}{@{}lrr@{}}
\toprule
\textbf{Recall@k} & \textbf{PatchFinder} & \textbf{Prospector} \\ 
\midrule
\textbf{K=1}      & 22.12\%              & 75.61\%             \\
\textbf{K=2}      & 25.42\%              & 80.42\%             \\
\textbf{K=3}      & 33.79\%              & 83.39\%             \\
\textbf{K=4}      & 39.2\%               & 85.52\%             \\
\textbf{K=5}      & 45.61\%              & 86.18\%             \\
\textbf{K=6}      & 48.29\%              & 86.93\%             \\
\textbf{K=7}      & 50.91\%              & 87.25\%             \\
\textbf{K=8}      & 52.14\%              & 87.54\%             \\
\textbf{K=9}      & 54.36\%              & 87.92\%             \\
\textbf{K=10}     & 55.72\%              & 88.16\%            \\
\bottomrule
\end{tabular}
\label{tab:recallatk}
\end{table*}

\subsubsection{On the \texttt{After-March-2024} dataset.}
Table \ref{tab:recallatk} shows the performance of Prospector and PatchFinder on the dataset.
Across all K values from 1 to 10, Prospector consistently outperforms PatchFinder by a significant margin. 
For example, at K=1, Prospector achieves a Recall@1 of 75.61\%, compared to PatchFinder's 22.12\%, indicating that Prospector is far more effective at identifying the correct VFC in the top-ranked result. Similarly, at K=10, Prospector achieves a Recall@10 of 88.16\%, compared to PatchFinder's 55.72\%, maintaining its superior performance as K increases. The steady improvement in recall scores for both methods as K increases reflects the benefit of considering a broader range of candidate VFCs. 
These results demonstrate that Prospector is more robust and accurate in tracing VFCs than PatchFinder, which is due to the difference between VFCs' considerations from the two tools. 
Prospector prioritizes commits that clearly link to vulnerability fixes, especially those mentioning vulnerability IDs in messages or referenced in advisories. 
It also values commits with bug-tracking or GitHub issue IDs from advisories, as they provide direct, verifiable links to vulnerabilities. These commits are considered the most reliable for identifying VFCs. 
Less important are commits with keywords from NVD descriptions or changes to files/methods mentioned in NVD, which add supporting context but are not as directly linked.
PatchFinder, on the other hand, uses deep learning without specific filters to exclude false candidates, which may weaken its ability to connect NVD records to their true VFCs accurately.
Furthermore, the observed decline in PatchFinder's performance may be attributable to the dataset difference.
The VFCs in both the fine-tuning dataset and the \texttt{After-March-2024} dataset are limited to patch-tagged Git references from NVD records, diverging from the training and evaluation data utilized in PatchFinder's original paper.
Relying on a learning-based framework trained with a random sample of 5,000 non-patch commits as negative examples, PatchFinder's performance may suffer from inconsistencies in training data. 
This limitation could hinder its adaptability across varied vulnerability contexts, exposing a fundamental challenge in refining such methods for this task.

\begin{figure*}[h]
\centering
\includegraphics[width=0.40\columnwidth]{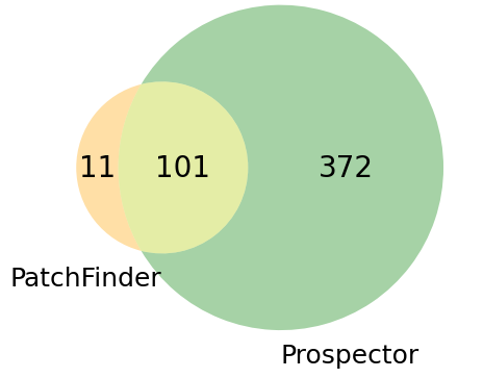} 
\caption{Venn diagram on the results of PatchFinder and Prospector on the \texttt{After-March-2024} dataset}
\label{fig:venn_rq4}
\end{figure*}

The Venn diagram in Figure ~\ref{fig:venn_rq4} presents another perspective of the comparison between PatchFinder and Prospector, illustrating the overlap between their Top-1 predictions.
As shown, Prospector successfully identifies VFCs for 473 NVD records, significantly outperforming PatchFinder, which identifies VFCs for only 112 NVD records. 
Additionally, we observe a significant overlap of 101 NVD records between the two tools, indicating that Prospector also covers the majority of NVD records identified by PatchFinder. 
Notably, Prospector identifies 372 unique NVD records with true VFCs, whereas PatchFinder identifies only 11 unique NVD records, highlighting a substantial difference in coverage. 
This suggests that heuristic approaches, such as Prospector, are currently more reliable and robust for VFC identification from Git-hosting repositories, particularly when handling large-scale, real-world datasets.  

\subsubsection{On the \texttt{Implicit VFCs} dataset.}
We then evaluate Prospector on implicit VFCs to determine whether it can detect \textit{more challenging} VFCs that have not yet been recorded in NVD. 
We did not evaluate the effectiveness of PatchFinder on the implicit VFC dataset due to the high computational cost of running PatchFinder and its limited effectiveness on the \texttt{After-March-2024} dataset.

\begin{table*}[h]
\caption{Prospector results on both \texttt{After-March-2024} explicit dataset and implicit dataset.}
\label{tab:prospector}
\begin{tabular}{@{}lrrrr@{}}
\toprule
& \textbf{\# Records} & \textbf{\# True Records} & \textbf{\# VFCs} & \textbf{\# True VFCs} \\ 
\midrule
\texttt{After-March-2024} & 510 & 473 (92.75\%) & 597  & 473 (79.2\%) \\
\texttt{Implicit VFCs} & 300 & 220 (73.3\%) & 322  & 235 (73.0\%) \\ 
\bottomrule
\end{tabular}
\begin{tablenotes}
\small
\item \textit{Note:}  ``\# Records'' and ``\# True Records'' denote the number of NVD records found with VFCs by Prospector and true NVD records found with VFCs by Prospector. ``\# VFC'' and ``\# True VFC'' denote the number of candidate and true VFCs identified by Prospector.
\end{tablenotes}
\end{table*}

Table~\ref{tab:prospector} shows the results of Prospector. 
Prospector identifies 597 candidate VFCs with a precision of nearly 80\% for explicit NVD records (on the \texttt{After-March-2024} dataset).
It also successfully associates these VFCs with 473 true NVD records, resulting in a precision of 92.75\% in terms of NVD records. 
However, the precision drops on the \texttt{Implicit VFCs} dataset, with 73.0\% for VFCs precision and 73.3\% for NVD records precision. 
These results suggest that implicit VFCs are more challenging to identify than explicit VFCs when using Prospector.

\begin{table}[h]
\centering
\caption{Results of Prospector running on a larger implicit dataset.}
\begin{tabular}{@{}lrr@{}}
\toprule
\textbf{\begin{tabular}[c]{@{}c@{}}\#NVD records \\ processed\end{tabular}} & \textbf{\begin{tabular}[c]{@{}c@{}}\#NVD records \\ found with VFCs\end{tabular}} & \textbf{\#VFCs} \\ 
\midrule
99,345 & 2,795 & 3,686  \\
\bottomrule
\end{tabular}
\label{tab:prospector_implicit}
\end{table}

\subsubsection{Automated Extraction Result.} Given the precision of Prospector on our \texttt{Implicit VFCs} dataset is over 70\%, we decide to run this method on the whole NVD records that do not have an explicit VFC mentioned in the reference, i.e., do not have a patch-tagged Git reference.
Out of the total 99,345 NVD records with public GitHub repositories links processed, only 2,795 records are found to have associated VFCs that have a Prospector score over 60. This resulted in a total of 3,686 high-confidence VFCs.

\find{
\textbf{Answer to RQ4:} 
    On our \texttt{After-March-2024} dataset, Prospector significantly outperforms PatchFinder in identifying unseen explicit VFCs from GitHub repositories.
    On our \texttt{Implicit VFCs} dataset, Prospector experiences a precision drop, from 79.2\% and 92.7\% on the \texttt{After-March-2024} dataset to around 73\% for both VFCs and NVD records, respectively. 
    These findings highlight that implicit VFCs are significantly more challenging to identify than explicit ones.
}

\subsection{RQ5: Comparison Among Different Sources}

% Our merged implicit VFC dataset contains a total of 9,394 NVD records and 17,299 VFCs, with a precision of 90\% and 87.9\%, respectively. 
% This dataset nearly doubles the size of all existing datasets and expands our explicit VFC dataset by almost 60\% and 74\% in terms of NVD records and VFCs. 

\begin{table}[h]
\centering
\caption{Statistics of our dataset by combining all the Steps.}
\label{tab:combined_dataset_final}
\begin{tabular}{@{}lrrrr@{}}
\toprule
\textbf{Dataset} & \textbf{\# Records} & \textbf{\# VFCs} & \textbf{$\text{SuccessRate}_{\text{Records}}$} & \textbf{$\text{Precision}_{\text{VFCs}}$}\\ 
\midrule
S1: From NVD Reference & 20,360 & 31,942     & 89.4\% & 87.0\% \\
S2: From External Databases   & 18,985    &  29,254  & 90.6\%  & 88.4\% \\ 
S3: From GitHub Repositories & 2,795    &  3,686  & 73.3\%  & 73.0\% \\
\midrule 
Full Dataset & 26,710  &   37,441   & 87.8\%       & 86.1\%  \\ 
\bottomrule
\end{tabular}
\end{table}

To address RQ5, we compare the effectiveness of the three primary sources used in this study - NVD references (S1), external security databases (S2), and GitHub repositories (S3) - in terms of precision, coverage, and overlap.
This analysis synthesizes the results from RQ2, RQ3, and RQ4 to provide a holistic view of how these methods contribute to mapping NVD records to their VFCs. 
% By combining the VFCs identified from each source, we aim to understand their complementary strengths and limitations, offering insights into their collective utility for constructing comprehensive vulnerability datasets.

\subsubsection{Precision} We evaluate \textit{precision} from two perspectives: $\text{SuccessRate}_{\text{Records}}$ (the percentage of sampled NVD records mapped to at least one true VFC) and $\text{Precision}_{\text{VFCs}}$ (the percentage of identified candidate VFCs confirmed as true VFCs). 
Table~\ref{tab:combined_dataset_final} summarizes these metrics for each source and the combined dataset.

\begin{itemize}[leftmargin=0pt, itemindent=*]
\item S1: NVD References: This source yields a $\text{SuccessRate}_{\text{Records}}$ of 89.4\% and a $\text{Precision}_{\text{VFCs}}$ of 87.0\% across 20,360 records and 31,942 VFCs. The high precision reflects the reliability of Git references, particularly those tagged with \texttt{Patch}, which provide direct links to VFCs. However, the automated pipeline's limitation to a two-level reference tree may exclude some valid VFCs from non-Git references, slightly hindering its precision.
\item S2: External Databases: Extracting VFCs from six external security databases (e.g., Snyk, GitHub Advisory) results in a $\text{SuccessRate}_{\text{Records}}$ of 90.6\% and a $\text{Precision}_{\text{VFCs}}$ of 88.4\% for 18,985 records and 29,254 VFCs. This source achieves the highest precision, likely due to the curated nature of these databases, which often include detailed patch information not present in the NVD.
\item S3: GitHub Repositories: Using Prospector on the implicit VFC dataset, this source achieves a $\text{SuccessRate}_{\text{Records}}$ of 73.3\% and a $\text{Precision}_{\text{VFCs}}$ of 73.0\% across 2,795 records and 3,686 VFCs. The lower precision compared to S1 and S2 highlights the challenge of identifying implicit VFCs, where heuristic-based methods struggle with less explicit links to vulnerabilities.
\end{itemize}

The full dataset, merging all three sources, achieves an overall $\text{SuccessRate}_{\text{Records}}$ of 87.8\% and $\text{Precision}_{\text{VFCs}}$ of 86.1\%. 
This weighted average reflects the dominant contribution of S1 and S2, tempered by the lower precision of S3 due to its smaller scale and focus on harder-to-detect implicit VFCs.

\subsubsection{Coverage} We define \textit{coverage} as the proportion of NVD records successfully mapped to at least one VFC relative to the total NVD dataset.
Table~\ref{tab:combined_dataset_final} shows the individual contributions:

\begin{itemize}[leftmargin=0pt, itemindent=*]
    \item S1: NVD References: Maps 20,360 unique NVD records, covering approximately 8.7\% of the NVD. This is the largest single-source contribution, driven by the automated pipeline's ability to extract VFCs from Git references and limited non-Git references.
    \item S2: External Databases: Identifies VFCs for 18,985 records, covering about 8.1\% of the NVD. While slightly lower than S1, this source complements NVD references by capturing VFCs absent from the NVD, especially from well-maintained databases like Snyk and GitHub Advisory.
    \item S3: GitHub Repositories: Contributes only 2,795 records (1.2\% of the NVD), reflecting the computational constraints and focus on implicit VFCs not found in S1 or S2. Despite its limited scale, S3 adds unique mappings that enhance overall coverage.
\end{itemize}

Combining all sources, the full dataset maps 26,710 unique NVD records to 37,441 VFCs from 7,634 projects, achieving a total coverage of approximately 11.3\% of the NVD.
This combined effort exceeds the 8.7\% coverage from NVD references alone, demonstrating the value of integrating multiple sources. 
However, with 88.7\% of NVD records still unmapped, substantial gaps remain, particularly for records lacking Git references (Categories 3 and 4).

\subsubsection{Overlap} 
\begin{figure}[ht]
    \centering
    \begin{subfigure}[t]{0.45\textwidth}
        \centering
        \includegraphics[width=\linewidth]{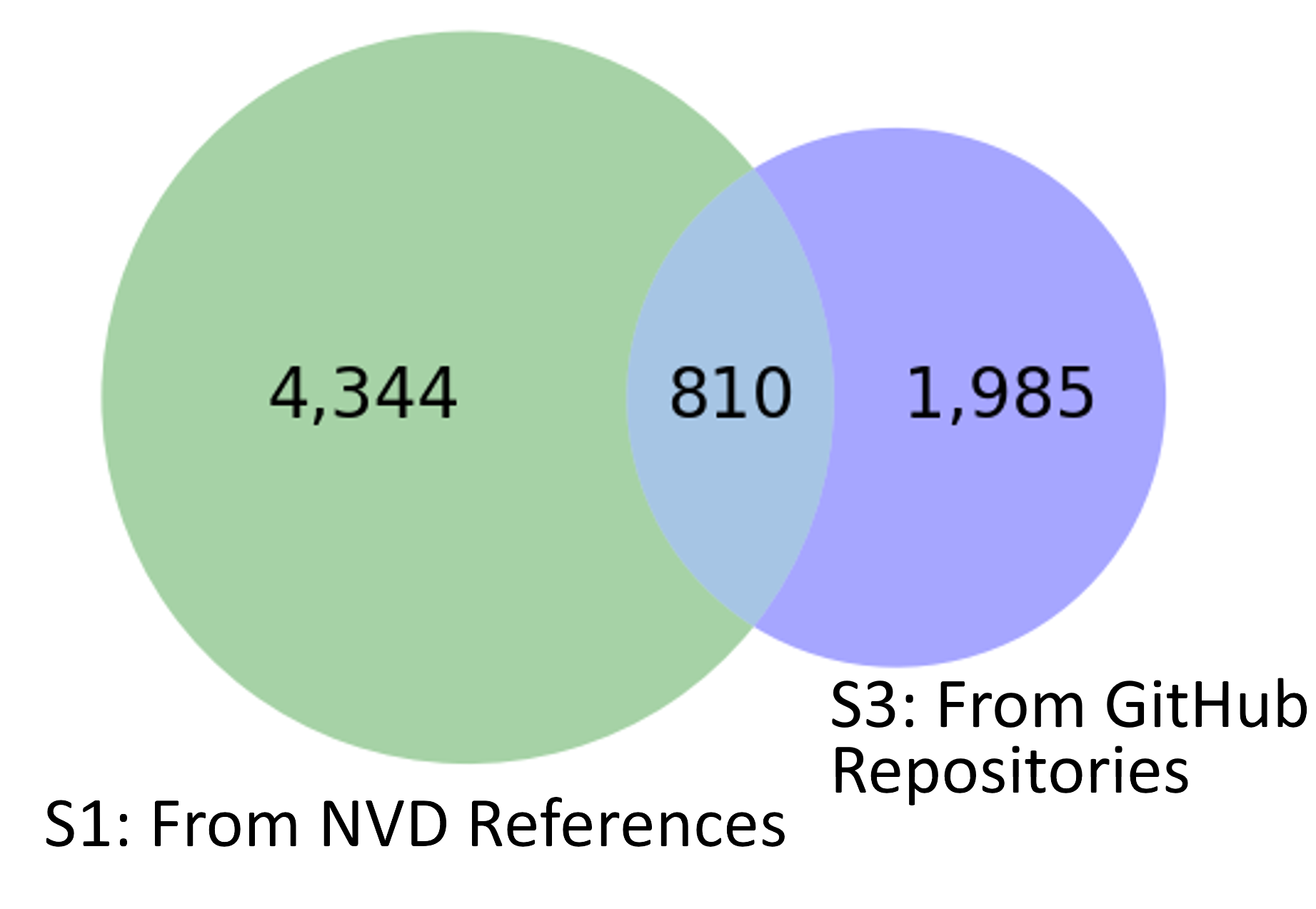}
        \caption{}
        \label{fig:venn_rq5_s1_s3}
    \end{subfigure}
    \hfill
    \begin{subfigure}[t]{0.43\textwidth}
        \centering
        \includegraphics[width=\linewidth]{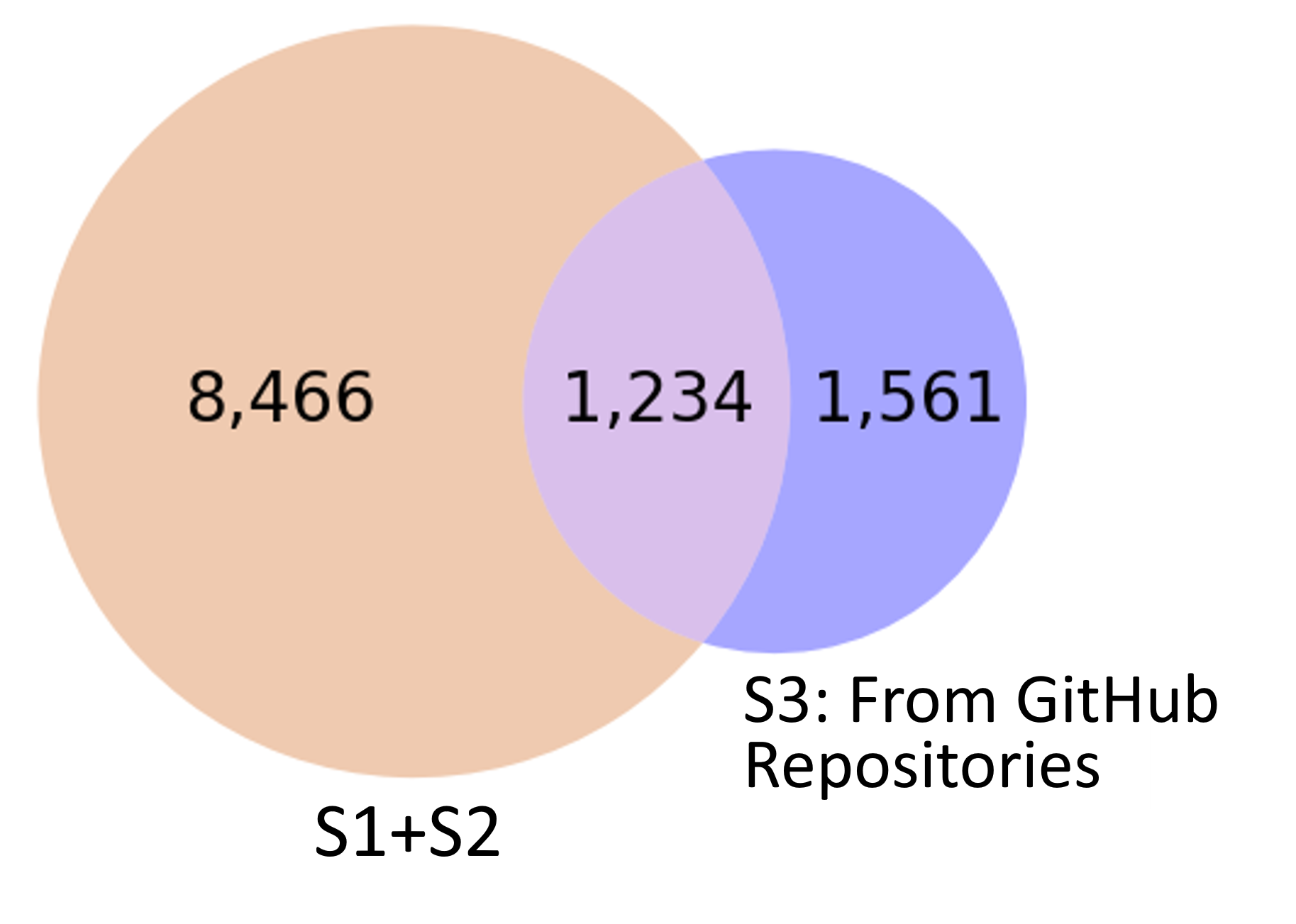}
        \caption{}
        \label{fig:venn_rq5_s12_s3}
    \end{subfigure}    
    \caption{Venn diagrams comparing at least one VFC found in NVD records by S1 vs. S3 (a) and S1+S2 vs. S3 (b)}
    \label{fig:venn_rq5}
\end{figure}

To assess redundancy and uniqueness, we analyze the overlap of NVD records and VFCs across the three sources.
In Section~\ref{sec:comparing_NVD_external}, we already compare the VFCs extracted from NVD references (S1) and external databases (S2), revealing significant overlap (e.g., 14,375 shared NVD records and 20,907 shared VFCs as shown in Figure~\ref{fig:venn_rq2}). 
Here, we extend this analysis to include S3, focusing on pairwise comparisons - S1 versus S3 and the combined S1+S2 versus S3 - to assess how these sources complement one another in the merged dataset.

\begin{itemize}[leftmargin=0pt, itemindent=*]
    \item S1 vs. S3: Figure~\ref{fig:venn_rq5_s1_s3} shows 810 NVD records common between NVD references and GitHub repositories, with 4,344 unique to S1 and 1,985 unique to S3. This moderate overlap suggests that the GitHub repository search complements the NVD references, and each source contributes unique VFCs, enhancing overall diversity.
    \item S1+S2 vs. S3: Figure~\ref{fig:venn_rq5_s12_s3} shows that 1,234 NVD records shared by S1+S2 (NVD reference and external security databases) and S3. This indicates that even by combining NVD references and external security databases, the GitHub repository search can still provide VFCs for unique NVD records.
\end{itemize}

The merged dataset of 26,710 records indicates that combining S1 and S2 captures the majority of mappings, with S3 providing a smaller but critical supplement. 
The overlap between S1 and S2 reduces the net gain from external databases, while S3's unique additions highlight the potential of repository-based methods for hard-to-find VFCs.

\subsubsection{Composition of Mapped NVD Records by Product Type}
\label{sec:cpe_composition}

Finally, to better understand the landscape of successfully mapped vulnerabilities, we analyzed the composition of our final dataset of 26,710 records by product type. We categorized each record according to the official product types defined in the CPE 2.3 specification: \texttt{a} for \textbf{applications}, \texttt{o} for \textbf{operating systems}, and \texttt{h} for \textbf{hardware}.

The results, summarized in Table~\ref{tab:cpe_composition}, reveal that our mapped dataset is overwhelmingly composed of application-level vulnerabilities. 
This heavily skewed distribution provides an interesting insight into where current mapping techniques are most effective. The dominance of applications strongly suggests that the prevalence of open-source software, public version control systems like Git, and more transparent disclosure practices make their VFCs significantly more accessible. In contrast, the low representation of OS and hardware vulnerabilities underscores the difficulty in mapping VFCs for products that often rely on proprietary codebases and distribute patches as binary files rather than source code commits.

\begin{table}[h!]
\centering
\caption{Numbers of Mapped NVD Records by CPE Product Type}
\label{tab:cpe_composition}
\begin{tabular}{@{}lrr@{}}
\toprule
\textbf{Product Type}        & \textbf{\# Records} & \textbf{Percentage} \\ \midrule
Applications (part:a)      & 24,583                     & 92.1\%                       \\ \midrule
Operating Systems (part:o) & 2,115                      & 7.9\%                        \\ \midrule
Hardware (part:h)          & 12                         & \textless 0.1\%              \\ \midrule
\textbf{Total}               & \textbf{26,710}            & \textbf{100\%}              
\end{tabular}%
\end{table}

\find{
\textbf{Answer to RQ5:} 
By leveraging NVD references, external databases, and GitHub repositories, we mapped 26,710 NVD records to 37,441 VFCs, achieving an overall $\text{SuccessRate}_{\text{Records}}$ of 87.8\% and $\text{Precision}_{\text{VFCs}}$ of 86.1\%. 
External databases (S2) provide the highest precision (90.6\% for records), followed closely by NVD references (89.4\%), while GitHub repositories (73.3\%) lag due to the complexity of implicit VFCs. 
Coverage reaches 11.3\% of the NVD, with NVD references contributing the largest share (8.7\%), supplemented by external databases (8.1\%) and GitHub repositories (1.2\%). 
Overlap between NVD references and external databases is significant, but GitHub repositories add unique mappings, enhancing the dataset's comprehensiveness.
}

\section{Discussion}
\label{sec:discussion}

\subsection{Lesson Learned}

\textbf{Lesson \#1:  VFCs from Git references are not enough for constructing a vulnerability dataset.} Our study reveals that VFCs identified through Git references come from 18,545 NVD records, which constitute only 7.9\% of the total NVD records in the NVD dataset. This highlights the inadequacy of relying solely on VFCs from Git references for constructing vulnerability datasets, resulting in a small-scale dataset. Such limited datasets may introduce bias in vulnerability research and lead to poor performance in learning-based models.

\vspace{4px}
\noindent\textbf{Lesson \#2: VFCs from NVD records without Git References should be mined to expand vulnerability datasets.}
In this study, we explore various methods to identify VFCs through manual investigation of related resources, such as NVD references, external databases, and GitHub repositories. Step 1 in our automated pipeline successfully identifies 29,249 VFCs from 18,545 NVD records with Git references. However, for NVD records without Git references, we were only able to identify 2,693 VFCs from 1,815 NVD records.
It is noteworthy that although NVD records without Git references constitute the majority in the NVD database, the VFCs we identified from these records represent only around 0.77\% of all NVD records, highlighting a substantial gap and room for improvement.
We recommend that future research prioritize mining VFCs from non-Git references to significantly enhance the number of vulnerable samples derived from the NVD. 
Broadening the
scope beyond Git references will not only enrich the comprehensiveness of vulnerability datasets but also improve the effectiveness and robustness of vulnerability detection and analysis tools that depend on NVD data.

% This highlights the need to mine implicit VFCs from various sources to significantly expand vulnerability datasets. By incorporating these implicit VFCs, we can obtain a more comprehensive and robust representation of vulnerabilities, which will improve the accuracy and performance of vulnerability detection and analysis tools.

% This highlights the inadequacy of relying solely on explicit VFCs for constructing vulnerability dataset. The limited coverage suggests that a vast majority of VFCs, from the implicit NVD records, remain undiscovered using current methodologies, potentially leading to incomplete vulnerability datasets. As a consequence, existing works on vulnerability and security that utilize these vulnerability datasets, which mostly consist of explicit VFCs only, can suffer the overfitting problem, as models trained on explicit VFCs may not generalize well to the broader set of vulnerabilities.
% We recommend that future research should prioritize the identification of implicit VFCs to significantly increase the number of vulnerable samples derived from the NVD. Expanding beyond explicit references will enhance the comprehensiveness and robustness of datasets, thereby improving the overall performance of vulnerability detection and analysis tools that rely on NVD data.

\vspace{4px}
\noindent\textbf{Lesson \#3: Learning-based methods for VFC identification from NVD are not ready yet}
Considerable effort has been invested in developing learning-based methods, utilizing either deep learning or LLMs, to detect VFCs within GitHub repositories. 
However, our evaluation reveals that PatchFinder, a leading state-of-the-art tool, delivers suboptimal performance on our dataset. 
Even after fine-tuning, PatchFinder achieves a Recall@1 of 22.12\%. Furthermore, it is outperformed by Prospector, a purely heuristic-based approach, on the same dataset.

One plausible explanation for this disparity lies in the differences between the datasets. 
First, both the training and evaluation data utilized in our study differ from the ones used by PatchFinder.
In addition, PatchFinder employs a random sampling of 5,000 non-patch commits as negative samples during its training. 
In our study, we fine-tuned PatchFinder using our explicit data before March, 2024, and similarly sampled 5,000 non-patch commits. 
Variations in the training data, combined with the non-heuristic nature of deep learning, may contribute to the observed inconsistencies in performance. 
These results underscore the necessity for further research and enhancement of learning-based techniques to advance the accuracy and reliability of VFC identification.

\vspace{4px}
\noindent\textbf{Lesson \#4: Heuristic-based methods still face challenges in handling implicit VFCs.} While Prospector significantly outperforms PatchFinder in identifying explicit VFCs on our \texttt{After-March-2024} dataset, our experimental results indicate a notable drop in precision for the \texttt{Implicit VFCs} dataset. 
Specifically, it achieves only around 73\% precision on both NVD records and VFCs. Although this precision is encouraging, we found it insufficient for reliably identifying implicit VFCs. Moreover, the significant execution times of several minutes per CVE also present practical challenges when handling over 200k NVD records in the NVD.
In addition, we are only able to find VFCs for 2,795 NVD records, which is 1.2\% of the total NVD records.
This indicates that the prospector has a low recall rate when identifying VFCs.

\subsection{Suggestions for Future Research}
Based on the challenges and lessons identified in our study, we propose the following actionable suggestions to advance the state of the art in mapping NVD records to their VFCs:

\textbf{Action \#1: Enhance the VFC identification from non-Git references.}
Our study found that the vast majority of NVD records lack direct Git references, and mining their non-Git counterparts is challenging. 
The feasibility of this direction is directly supported by our RQ1 manual investigation results, which serve as supporting evidence that VFCs can be retrieved from non-Git references. Human annotators successfully identified true VFCs in 17\% of Category 3 and 9.9\% of Category 4 records by following reference links and interpreting contextual information. This confirms that the required information is present and retrievable; the proposed advanced techniques therefore aim to automate a process humans have already demonstrated is possible. To bridge this gap, future work should focus on more intelligent data extraction techniques. 

Firstly, instead of relying on simple hyperlink traversal, researchers could build Natural Language Processing (NLP) models to parse the content of linked security advisories, bug reports, and mailing list archives. These models could be trained to recognize and extract entities like commit hashes, repository URLs, patch file names, and version numbers, even when they are embedded in unstructured text.
Secondly, automated tools could be designed to navigate complex, domain-specific websites like Bugzilla or project-specific issue trackers. This would involve creating scrapers that can handle unique page layouts and workflows to follow discussion threads that eventually lead to a VFC.

\textbf{Action \#2: Develop Hybrid VFC Identification Approaches.}
Our results show that both purely heuristic-based tools (like Prospector) and learning-based methods (such as PatchFinder) have distinct limitations. While heuristics are more reliable for explicit VFCs, they struggle with recall for implicit cases; conversely, learning-based models show promise, but are not yet robust enough for general use. 
It is worth noting that our pipeline deliberately applies crawler-based extraction for cases where Git links are directly available (RQ2 and RQ3), reserving more advanced methods for harder implicit cases (RQ4). This is a rational design choice: advanced learning-based methods such as PatchFinder require approximately 28 GPU hours for just 510 NVD records, making them computationally prohibitive at the scale of the full NVD corpus, whereas our crawler-based pipeline achieves over 87\% precision across 20,360 records, in only around 25 hours. That said, our RQ4 results reveal that neither heuristic nor learning-based methods alone are sufficient for implicit VFCs, motivating a hybrid design. The feasibility of such an approach is grounded in the complementary performance observed: Prospector achieves high precision on explicit VFCs but drops notably on implicit ones, while learned representations can capture semantic relationships that heuristics miss. A hybrid pipeline is feasible because the two methods fail in different ways and can therefore compensate for each other. Heuristics are fast and precise, but miss many implicit VFCs; learning-based models are better at capturing subtle semantic relationships but are too expensive to run over the entire NVD.

A promising direction is to create hybrid approaches that combine their respective strengths in a multi-stage pipeline. The first stage could use a refined heuristic-based filter to rapidly scan a repository and identify a small set of high-probability VFC candidates. This filter could be enhanced with deeper code analysis, such as parsing a commit’s Abstract Syntax Tree (AST) to better align code changes with the vulnerability type. The second stage would then apply a sophisticated learning-based model to re-rank only this small candidate set. This model could be a multi-modal architecture trained on high-quality data, including "hard negatives" which are non-VFCs with commit messages closely related to NVD descriptions (e.g., general bug fixes in the same vulnerable files), to deeply analyze the semantic relationship between the CVE description and the candidate commit, overcoming the semantic gap issue.
This hybrid strategy would leverage the speed and precision of heuristics for broad-scale filtering while reserving the powerful (but computationally expensive) semantic understanding of learning models for the most promising candidates.

\textbf{Action \#3: Develop Semi-Automated VFC Verification Techniques.} 
Our study's heavy reliance on manual verification, while rigorous, is labor-intensive and presents a scalability challenge. 
Specifically, the manual verification of 3,627 candidate VFCs sampled across our evaluations required an estimated 105 person-hours of focused expert effort. 
To improve the scalability of creating large, high-quality datasets, future work should focus on semi-automated verification methods.

To validate this direction, we conducted a preliminary experiment using OpenAI’s gpt-oss-20b model~\cite{gpt_oss} to verify a sample of 200 candidate VFCs from NVD records from Table~\ref{tab:4_2_result}. To construct the sample, we selected 100 CVEs, with one true VFC and one false VFC for each, resulting in a balanced dataset of 200 candidates (100 true VFCs and 100 false VFCs) to ensure a fair evaluation of the model's capabilities. We provided the model with the CVE details, commit message, and code diff for each candidate. The results are promising. On performance, from the 200 candidates, the model correctly identified 72 true positives, 89 true negatives, 11 false positives and 28 false negatives. This results in 86.7\% precision and 72\% recall. The strong precision is particularly valuable, as it suggests LLMs can be used to reliably filter and confirm true VFCs, significantly reducing the manual workload. On efficiency, the entire process for all 200 candidates, which was done with 1 NVIDIA H100 GPU (80GB RAM), took 3,617 seconds, averaging about 18.1 seconds per candidate. This represents a large improvement in time efficiency compared to the several minutes required for each human validation. 
Our analysis of the 28 false negatives suggests the model's recall is limited when crucial context beyond the commit and CVE description is unavailable. Some of these cases occurred with "silent VFCs"—where developers fix a vulnerability with ambiguous commit messages, deliberately omitting security-specific details or internal tracking references to prevent alerting potential attackers to exploitable vulnerabilities~\cite{zhou2021finding, sun2023silent, zhou2023colefunda}. Another reason for these false negative cases is that the deciding evidence is located in external artifacts like issue tracker links, pull requests, or separate security reports, which current LLMs are not able to capture. This suggests that while LLMs are powerful assistants, their accuracy is constrained by the input data, and they are not yet a full replacement for human expertise in complex cases.

\subsection{Threats to Validity}
\subsubsection{Threats to Internal Validity}
The manual investigation and verification in RQ1-RQ4 may introduce biases, as annotators might overlook certain VFCs due to methodological limitations or variability in interpretation. 
Additionally, ensuring the correctness of all identified VFCs is challenging, as it requires expertise across diverse vulnerability contexts.
To mitigate these threats, annotators reviewed a large number of vulnerabilities to gain domain knowledge before labeling.

\subsubsection{Threats to External Validity}
The evaluation of automated tools like Prospector and PatchFinder was limited to the \texttt{After-March-2024} dataset due to computational constraints. 
For example, preparing PatchFinder's dataset required approximately 10GB of storage and 28 hours of GPU time, making it impractical to scale to the entire NVD.
Our results may not be generalized to other datasets.
To mitigate this, we included all the explicit VFCs identified after March 2024 until September 2024.
This evaluation dataset reflects varying repository structures, programming languages, and vulnerability types, ensuring the evaluation remains broadly representative.

\subsubsection{Threats to Construct Validity}
The selection of evaluation metrics could potentially introduce threats to construct validity in our study. 
To address this, we have employed a varied set of metrics tailored to different RQs. To assess whether a method accurately identifies VFCs, we evaluate precision from two distinct perspectives: the NVD record level and the VFC level. 
This dual approach offers a more holistic view of each method's performance. 
Given that a single NVD record may correspond to multiple VFCs, calculating recall proves challenging and non-trivial. 
Consequently, we consider precision a suitable metric, effectively reflecting the performance capabilities of each method.
% The validity of the two metrics Precision\(_{Records}\) and Precision\(_{VFCs}\) is potentially affected by some threats. 
% For Precision\(_{Records}\), biased sampling of NVD records and incomplete identification of true VFCs may misrepresent precision, while reliance on inconsistent or incomplete NVD references can reduce accuracy. 
% For Precision\(_{VFCs}\), ambiguous candidate definitions, human error during manual verification, and limitations in heuristic filtering may result in false positives or overlooked valid commits. 
% Both metrics are further affected by tool-specific biases, repository structures, and the evolving nature of the NVD dataset. 

\section{Related Work}
\label{sec:related_work}
In addition to the studies discussed in Section~\ref{sec:mapping} that focus on mapping NVD or CVE records to their VFCs, other research has centered on identifying VFCs~\cite{zhou2017automated, nguyen2022vulcurator, nguyen2023multi, patchscout, wang2022vcmatch, xu2022tracking,irsan2026revisiting}. 
These methods can be broadly categorized into two classes: VFC identification and silent VFC identification~\cite{sun2023silent}, depending on whether commit messages are explicitly utilized.

For VFC identification, tools like VulCurator~\cite{nguyen2022vulcurator} and Hermes~\cite{nguyen2022hermes} leverage various information sources, such as commit messages and GitHub issues, to classify VFCs. 
Recently, Bui et al.~\cite{bui2024javavfc} proposed a large-scale VFC dataset from Java OSS. They manually annotated 784 VFCs and developed a set of keywords for filtering VFCs based on commit messages. Using this automated keyword-based filtering, their dataset expanded to include 16,837 VFCs.

Silent VFCs refer to VFCs without explicit log messages indicating the fixed vulnerability. Recent studies have introduced methods to identify these silent VFCs~\cite{zhou2021finding, zhou2023colefunda, nguyen2023multi,cheng2025fixseeker}. 
For instance, VulFixMiner~\cite{zhou2021finding} is the first deep-learning-based approach for detecting silent VFCs. 
It fine-tunes CodeBERT on cross-project and cross-language data to generate contextual embedding vectors for each commit based on code changes in affected files. These embeddings are then used to compute a prediction score, representing the likelihood that a commit addresses a vulnerability.

While these approaches advance VFC detection, they prioritize identification over tracing VFCs back to NVD records. 
Consequently, their applicability to our objective, mapping NVD records to their corresponding VFCs, is limited. 
Our work complements these identification-focused studies by addressing this mapping task, offering a broader perspective on VFC research.
% \rw{Could we add one sentence about what we did in this paper as we cannot use the related works we have highlighted here?}

\section{Conclusion and Future Work}
\label{sec:conclusion}
In this work, we conducted a comprehensive exploratory study to examine how hard it is to map NVD records to their corresponding VFCs.
Our empirical study confirms that mapping NVD records to their VFCs is remarkably difficult. 
Across the sampled NVD records, our manual analysis achieved over 86\% success with Git references but only 13\% or less with non-Git ones.
Our automated pipeline extracted 31,942 VFCs from 20,360 records (87\% precision), external databases added 29,254 VFCs for 18,985 records (88.4\% precision), and Prospector on GitHub mapped 3,686 VFCs for 2,795 records (73\% precision). 
Despite these efforts, we linked just 26,710 records, 11.3\% of the whole NVD records, leaving  88.7\% unmapped, especially records with only non-Git references.

The core challenge lies in the NVD's reliance on sparse, explicit links: Git references cover only 7.9\%, and external sources and repository searches struggle with precision and scale for the rest. 
Yet, our 37,441 VFCs provide a valuable starting point for analysis and detection.

Looking ahead, we plan to further enhance automation to parse non-Git references (e.g., via deeper scraping or NLP) and enhance the precision of the GitHub repository analysis tool to improve VFC identification and dataset quality. 
Additionally, we aim to leverage this dataset to advance other areas of vulnerability-related research, such as automated vulnerability detection~\cite{li2025out,bui2025vulcoco} and repair~\cite{yang2025semantics}.

\section*{Availability}
The replication package is available at \url{https://github.com/hungkien05/VFC-from-NVD-study}.

\section*{Acknowledgement}
This research / project is supported by the National Research Foundation, Singapore and Ministry of Digital Development \& Information under its Smart Nation and Digital Government Translational R\&D Grant (Award No: TRANS2026-TGC01). Any opinions, findings and conclusions or recommendations expressed in this material are those of the author(s) and do not reflect the views of National Research Foundation, Singapore.

% We start with a manual study to understand how humans find VFCs by checking NVD references.
% Getting the insights from the manual investigation, we built an automated pipeline to extract VFCs from NVD references.
% To examine the quality, we conducted manual verification and found that over 80\% of the Git references identified were true VFCs, and 80\% of the Git-links extracted non-Git references were true VFCs.
% Second, we extended a recent methodology to search for VFCs within relevant GitHub repositories, uncovering additional VFCs.
% Third, we collected existing VFC information from external security databases. 
% Finally, we compared the VFCs we found from different strategies and found that the VFCs we can find are bounded by the Git-links in the reference section and the CPE information.

\newpage
\bibliographystyle{ACM-Reference-Format}
\bibliography{main}

\end{document}